\documentclass[12pt]{article}
\usepackage{fullpage}
\usepackage{pstricks,pst-coil,multido}
\usepackage{amsmath,amssymb,bm}

\psset{linewidth=0.5pt,coilaspect=35,coilwidth=0.1,dash=2pt 1.5pt}

\allowdisplaybreaks[1]
\def\be#1\ee{\begin{equation}#1\end{equation}}
\def\bal#1\eal{\begin{align}#1\end{align}}
\def\bmu#1\emu{\begin{multline}#1\end{multline}}
\def\bga#1\ega{\begin{gather}#1\end{gather}}
\newcommand{\ba}{\begin{array}}
\newcommand{\ea}{\end{array}}
\newcommand{\n}{\notag}

\newcommand{\abs}[1]{\lvert#1\rvert}
\renewcommand{\d}{\partial}
\renewcommand{\bf}{\mathbf}
\renewcommand{\cal}{\mathcal}

\newcommand{\ds}{\displaystyle}

\title{\textbf{Equal-time two-point correlation functions \\
in Coulomb gauge Yang-Mills theory}}

\author{D. Campagnari$^{\,a}$, 
A. Weber$^{\,b}$\thanks{Email: \texttt{axel@ifm.umich.mx}}\:\,, 
H. Reinhardt$^{\,a}$, 
F. Astorga$^{\,b}$\thanks{Email: \texttt{astorgadeita@gmail.com}}\:\,,\\
W. Schleifenbaum$^{\,a}$ \\[2mm]
\and
\normalsize ${}^a$Institut f\"ur Theoretische Physik, 
Universit\"at T\"ubingen,\\
\normalsize Auf der Morgenstelle 14, D-72076 T\"ubingen, Germany
\and
\normalsize ${}^b$Instituto de F\'{\i}sica y Matem\'aticas,
Universidad Michoacana de San Nicol\'as de Hidalgo\\
\normalsize Edificio C-3, Ciudad Universitaria,\\ 
\normalsize A. Postal 2-82, 58040 Morelia, Michoac\'an, Mexico}

\date{\normalsize \today}

\begin{document}

\maketitle

\begin{abstract}
We apply a functional perturbative approach to the calculation of
the equal-time two-point correlation functions and the potential between
static color charges to one-loop order in Coulomb gauge Yang-Mills 
theory. The functional approach proceeds through a solution 
of the Schr\"odinger equation for the vacuum wave functional to order 
$g^2$ and derives the equal-time correlation
functions from a functional integral representation via new diagrammatic
rules. We show that the results coincide with those obtained from the
usual Lagrangian functional integral approach, extract the beta function,
and determine the anomalous dimensions of the equal-time gluon
and ghost two-point functions and the static potential under the assumption of
multiplicative renormalizability to all orders.
\end{abstract}

\newpage
\section{Introduction}

Finding an accurate (semi-)analytical description of the infrared sector
of QCD is still one of the most important challenges of present-day
quantum field theory. In this work we concentrate on Yang-Mills theory,
QCD without dynamical quarks, since it is in this sector
where the peculiar properties of QCD, in particular the confining
interaction between quarks, arise. Recently, much of the activity in this
area has focused on the formulation and (approximate) solution
of Yang-Mills theory in the Coulomb
gauge \cite{SS01}--\cite{PR09}, the primary reason being that the Coulomb 
gauge Hamiltonian explicitly contains the color-Coulomb potential which 
furnishes the dominant nonperturbative contribution to the static or heavy 
quark potential.

Semi-analytical functional approaches to the calculation of gluon and ghost 
propagators in the infrared, mostly using Dyson-Schwinger equations, have
been successful in Landau gauge Yang-Mills theory \cite{SHA97}--\cite{FMP09}. 
In the so-called ghost dominance approximation, even very simple analytical
solutions exist in the far infrared \cite{SLR06}, \cite{Zwa02}--\cite{PLN04}. 
Although the consistency of these solutions is still under discussion, it is 
natural to inquire whether a similar approach could be useful in the Coulomb 
gauge. The breaking of Lorentz covariance through the Coulomb gauge 
condition makes the usual Lagrangian functional integral approach quite 
cumbersome in this gauge, see, e.g., Ref.\ \cite{WR07a}. For this reason,
semi-analytical approaches in Coulomb gauge have mostly used a Hamiltonian
formulation. A set of equations similar to Dyson-Schwinger equations is
obtained from a variational principle using a Gaussian type of ansatzes for 
the vacuum wave functional in the Schr\"odinger representation 
\cite{SS01}, \cite{FR04a}--\cite{ERS07}. In the ghost dominance approximation, 
furthermore, simple analytical solutions are available for the far infrared 
\cite{Zwa04,SLR06}.

Nevertheless, the status of the semi-analytical and analytical solutions in the
Coulomb gauge is not yet entirely clear, for two reasons: first, two different 
solutions with an infrared scaling behavior (differing in the infrared 
exponents) have been found in both the analytical and the semi-analytical 
approaches \cite{Zwa04}--\cite{SLR06}, and 
there is as yet no theoretical guidance to what the physical solution should 
be; second, the inclusion of the Coulomb form factor (the form factor for the 
color-Coulomb potential, which measures the deviation of the Coulomb potential 
from a factorization in terms of ghost propagators) in the set of equations
of Dyson-Schwinger type results problematic. In Refs.\ 
\cite{FR04a}--\cite{ERS07}, the equation for the Coulomb form factor has been 
considered subleading compared to the equations for the gluon and ghost 
propagators and therefore treated in the tree-level approximation, while in 
Ref.\ \cite{ERS08} all equations have been considered to be of the same order 
and therefore treated on an equal footing,
with the result that solutions with infrared scaling behavior cease to exist.
It should be emphasized that only solutions with scaling behavior can
give rise to a linearly rising Coulomb potential, and that the latest lattice
calculations also show a scaling behavior for the equal-time correlation
functions in the deep infrared \cite{BQR09,NVI09}. It is not clear at present
how to improve the approximation used in the variational approach in order
to arrive at a unique and consistent solution.

An interesting relation between Landau and Coulomb gauge 
Yang-Mills theory has been pointed out in the ghost dominance approximation 
in Refs.\ \cite{Zwa04,SLR06}: the equal-time correlation functions 
of the Hamiltonian approach in
Coulomb gauge are the formal counterparts in three dimensions of the
covariant correlation functions in Landau gauge in four dimensions.
Building on this analogy, a possible strategy would be to replace the
variational principle by a calculation of equal-time correlation functions
in the Coulomb gauge and intend to formulate Dyson-Schwinger equations for
the latter. In the present work, we take a first step in this direction:
we set up a functional integral representation of the equal-time correlation
functions (without taking a detour to the space-time correlation functions)
that is the precise three-dimensional analogue of the usual
functional integral representation of the covariant correlation functions
in the Lagrangian approach to Landau gauge Yang-Mills theory. We
also develop a diagrammatic representation and a set of Feynman rules for
the equal-time correlation functions. We use this formulation 
to calculate the equal-time gluon and ghost
two-point correlation functions and the potential for static color charges
in Coulomb gauge perturbatively to one-loop order. We extract the one-loop
beta function and determine the asymptotic ultraviolet behavior of the
equal-time two-point functions and the static potential. We also show that 
our results coincide with those obtained in a Lagrangian functional integral 
approach \cite{WR07b,WR08} and use the latter for the renormalization of the
equal-time correlation functions and the static potential.

The organization of the paper is as follows: in the next section,
we determine the vacuum wave functional perturbatively to order $g^2$ from 
the solution of the Schr\"odinger equation. With the vacuum functional
determined to the corresponding order, we turn to the calculation of
the equal-time gluon and ghost two-point correlation functions in Section 3.
We also calculate the one-loop corrections to the static or heavy quark
potential (and thus to the Coulomb form factor) in the same section. 
Although for the latter calculation we need to go beyond the terms
that we have calculated for the vacuum functional in Section 2, 
the relevant additional contributions are quite simply determined. In 
Section 4, we provide another representation of the equal-time 
two-point functions by choosing equal times (zero) in the 
space-time correlation functions determined before in 
the Lagrangian functional integral representation \cite{WR07b,WR08} of the 
theory. The static potential can also be obtained from a two-point function
that arises in the Lagrangian approach. We use the alternative representations
of the two-point functions and the static potential to perform the
renormalization of our results. We show the nonrenormalization of the
ghost-gluon vertex in the same section and use it to determine the beta
function and the asymptotic ultraviolet behavior of the two-point functions.
We also show that the same beta function is found from considering the
static potential. Finally, in Section 5, we 
summarize our findings and comment on several possible applications.
In the Appendix, we give some details on an important difference that arises 
between the Lagrangian and the Hamiltonian approach when it comes to the 
implementation of the Coulomb gauge.

\section{Perturbative vacuum functional}

It is very simple to write down a functional integral representation of the
equal-time correlation functions, given that they are nothing but the
vacuum expectation values of products of the field operators. In
the Schr\"odinger representation of Yang-Mills theory in Coulomb gauge, 
the equal-time $n$-point correlation functions in (3-)momentum space
have the following representation:
\bmu
  \langle A_i^a (\bf{p}_1, t=0) A_j^b (\bf{p}_2, t=0) \cdots 
  A_r^f (\bf{p}_n, t=0) \rangle \\
  = \int D [\bf{A} ] \, \delta( \nabla \cdot \bf{A} ) \, \textsf{FP} (\bf{A})
  \, A_i^a (\bf{p}_1) A_j^b (\bf{p}_2) \cdots A_r^f (\bf{p}_n) \,
  \abs{\psi (\bf{A})}^2 \:. \label{corrfunc}
\emu
Here, $\psi (\bf{A})$ is the true vacuum wave functional of the 
theory. The (absolute) square $\abs{\psi (\bf{A})}^2$ then plays the r\^ole of
the exponential of the negative Euclidean classical action 
in the corresponding representation of the covariant correlation
functions (in Euclidean space). $\textsf{FP} (\bf{A}) \equiv \text{det}
[-\nabla \cdot \bf{D} (\bf{A})]$, with the covariant derivative in 
the adjoint representation defined as
\be
\bf{D}^{ab} (\bf{A}) = \delta^{ab} \, \nabla + g f^{abc} \bf{A}^c \:, 
\ee
is the Faddeev-Popov determinant (in
3 dimensions) which forms a part of the integration measure for the scalar
product of states in the Schr\"odinger representation (see Ref.\ 
\cite{CL80}). Note that the fields $A_i^a (\bf{p})$ on the left-hand side
of Eq.\ \eqref{corrfunc} are spatially transverse, $\bf{p} \cdot \bf{A}^a
(\bf{p}) = 0$. We will assume the transversality of the fields 
$\bf{A}^a$ in all of the following formulae, which
we could make manifest by introducing a transverse basis in momentum
space. However, there is usually no need to do so explicitly.

In order to write down the functional integral for the equal-time
correlation functions explicitly, the vacuum wave functional
needs to be specified. The analogy with the covariant theory
suggests to use an exponential ansatz for this wave functional,
in the spirit of the $e^S$ expansion in many-body physics \cite{KLZ78}.
We consider a full Volterra expansion of the exponent:
\bmu
  \psi (\bf{A}) = \exp \bigg( - \sum_{k=2}^\infty \frac{1}{k!}
  \int \frac{d^3 p_1}{(2 \pi)^3} \cdots \frac{d^3 p_k}{(2 \pi)^3} \, 
  \sum_{i_1, i_2, \ldots, i_k} \, \sum_{a_1, a_2, \ldots, a_k} 
  f_{k; i_1 i_2 \ldots i_k}^{a_1 a_2 \ldots a_k} 
  (-\bf{p}_1, \ldots, -\bf{p}_k) \\
  \times A_{i_1}^{a_1} (\bf{p}_1) \cdots A_{i_k}^{a_k} (\bf{p}_k) 
  (2 \pi)^3 \delta (\bf{p}_1 + \ldots + \bf{p}_k) \bigg) \:. \label{vacans}
\emu
Any normalization factor can be conveniently absorbed in the
functional integration measure in Eq.\ \eqref{corrfunc}.
Terms linear in $\bf{A}$ in the exponent ($k=1$) are excluded by the symmetry 
of the wave functional under global gauge transformations (in the
absence of external color charges). Regarding notation, given
that our Hamiltonian formalism is not manifestly covariant, we will denote 
the contravariant spatial components of 4-vectors by (latin) 
\emph{subindices}.

We insert this ansatz for the vacuum wave functional into the Schr\"odinger 
equation 
\be
H \psi (\bf{A}) = E_0 \psi (\bf{A}) \:, \label{ETSE}
\ee 
where $H$ is the
Christ-Lee Hamiltonian for Coulomb gauge Yang-Mills theory \cite{CL80},
\bal
  H &= \frac{1}{2} \int d^3 x \left( - \frac{1}{\textsf{FP} (\bf{A})}
  \frac{\delta}{\delta A_i^a (\bf{x})} \textsf{FP} (\bf{A})
  \frac{\delta}{\delta A_i^a (\bf{x})} +
  B_i^a (\bf{x}) B_i^a (\bf{x}) \right) \n \\
  &\phantom{=} {}+ \frac{g^2}{2} \int d^3 x \, d^3 y \,
  \frac{1}{\textsf{FP} (\bf{A})} \, \rho^a (\bf{x}) \,
  \textsf{FP} (\bf{A}) \, \langle \bf{x}, a | (-\nabla \cdot \bf{D})^{-1}
  (-\nabla^2) (-\nabla \cdot \bf{D})^{-1} | \bf{y}, b \rangle  \,
  \rho^b (\bf{y}) \:. \label{christlee}
\eal
Here,
\be
  B^a_i = -\frac{1}{2} \, \epsilon_{i j k} F^a_{j k} =
  \left( \nabla \times \bf{A}^a -
  \frac{g}{2} \, f^{abc} \bf{A}^b \times 
  \bf{A}^c \right)_i
\ee
is the chromo-magnetic field, and
\be
  \rho^a (\bf{x}) = \rho^a_q (\bf{x}) + f^{abc} A_j^b (\bf{x}) \, 
  \frac{1}{i} \frac{\delta}{\delta A_j^c (\bf{x})} \label{colch}
\ee
the color charge density, including external static charges $\rho_q$
for later use. Note that we have extracted a factor $g$ from the
color charges in order to simplify the counting of orders of $g$ in the
rest of the paper. The notation $\langle \bf{x}, a | C | \bf{y}, b \rangle$
refers to the kernel of the operator $C$ in an integral representation.

In most of the following perturbative calculation, we will need the 
Hamiltonian only up to order $g^2$, where
\bal
  \lefteqn{H = \frac{1}{2} \int \frac{d^3 p}{(2 \pi)^3} \left( -(2 \pi)^3 
  \frac{\delta}{\delta A_i^a (\bf{p})} (2 \pi)^3
  \frac{\delta}{\delta A_i^a (-\bf{p})} + A_i^a (-\bf{p}) \, 
  \bf{p}^2 A_i^a (\bf{p}) \right)} \hspace{6mm} \label{clA2} \\
  &{}+ \frac{1}{2} \int \frac{d^3 p}{(2 \pi)^3} \,
  A_i^a (-\bf{p}) \left( \frac{N_c g^2}{2} \int \frac{d^3 q}{(2 \pi)^3}
  \frac{1 - (\hat{\bf{p}} \cdot \hat{\bf{q}})^2}{(\bf{p} - \bf{q})^2} \right)
  (2 \pi)^3 \frac{\delta}{\delta A_i^a (-\bf{p})} \label{cls1} \\
  &{}+ \frac{g}{3!} \int \frac{d^3 p_1}{(2 \pi)^3}
  \frac{d^3 p_2}{(2 \pi)^3} \frac{d^3 p_3}{(2 \pi)^3} \, i f^{abc} \left[
  \delta_{jk} (p_{1,l} - p_{2,l}) + \delta_{kl} (p_{2,j} - p_{3,j})
  + \delta_{lj} (p_{3,k} - p_{1,k}) \right] \n \\
  &\hspace{3cm} {}\times A^a_j (\bf{p}_1) A^b_k (\bf{p}_2) A^c_l (\bf{p}_3)
  (2 \pi)^3 \delta (\bf{p}_1 + \bf{p}_2 + \bf{p}_3) \label{clA3} \\
  &{}+ \frac{g^2}{4!} \int \frac{d^3 p_1}{(2 \pi)^3}
  \cdots \frac{d^3 p_4}{(2 \pi)^3} \left[ f^{abe} f^{cde} (\delta_{ik} 
  \delta_{jl} - \delta_{il} \delta_{jk}) + f^{ace} f^{bde} (\delta_{ij} 
  \delta_{kl} - \delta_{il} \delta_{jk}) \right. \n \\
  &\phantom{+} \left.
  {}+ f^{ade} f^{bce} (\delta_{ij} \delta_{kl} - \delta_{ik} \delta_{jl})
  \right] A^a_i (\bf{p}_1) A^b_j (\bf{p}_2) A^c_k (\bf{p}_3) A^d_l (\bf{p}_4) 
  (2 \pi)^3 \delta (\bf{p}_1 + \bf{p}_2 + \bf{p}_3 + \bf{p}_4) \label{clA4} \\
  &{}+ \frac{g^2}{2} \int \frac{d^3 p}{(2 \pi)^3} \,
  \rho^a (-\bf{p}) \frac{1}{\bf{p}^2} \, \rho^a (\bf{p}) + \cal{O} (g^3) \:.
  \label{clcoul}
\eal
The term \eqref{cls1} stems from the
application of the functional derivative to the Faddeev-Popov determinant.
In this term, $N_c$ stands for the number of colors,
$f^{acd} f^{bcd} = N_c \delta^{ab}$,
and $\hat{\bf{p}} \equiv \bf{p}/|\bf{p}|$ denotes a unit vector.
In the \emph{absence} of external charges, we get for the term \eqref{clcoul}
\bal
  \lefteqn{\frac{g^2}{2} \int \frac{d^3 p}{(2 \pi)^3} \,
  \rho^a (-\bf{p}) \frac{1}{\bf{p}^2} \, \rho^a (\bf{p})} \hspace{1cm} \n \\
  &= \frac{1}{2} \int \frac{d^3 p}{(2 \pi)^3} \,
  A_i^a (-\bf{p}) \left( N_c g^2 \int \frac{d^3 q}{(2 \pi)^3}
  \frac{t_{ij} (\bf{q})}{(\bf{p} - \bf{q})^2} \right)
  (2 \pi)^3 \frac{\delta}{\delta A_j^a (-\bf{p})} \label{cls2} \\
  &\phantom{=} {}- \frac{g^2}{4} \int \frac{d^3 p_1}{(2 \pi)^3}
  \cdots \frac{d^3 p_4}{(2 \pi)^3} \left( 
  f^{ace} f^{bde} \, \frac{\delta_{ik} \delta_{jl}}{(\bf{p}_1 + \bf{p}_3)^2} 
  + f^{ade} f^{bce} \, \frac{\delta_{il} \delta_{jk}}{(\bf{p}_1 + \bf{p}_4)^2} 
  \right) \n \\
  &\phantom{= {}-} {}\times (2 \pi)^3 \delta(\bf{p}_1 + \bf{p}_2 + \bf{p}_3 
  + \bf{p}_4) \, A^a_i (\bf{p}_1) A^b_j (\bf{p}_2)
  (2 \pi)^3 \frac{\delta}{\delta A_k^c (-\bf{p}_3)} (2 \pi)^3
  \frac{\delta}{\delta A_l^d (-\bf{p}_4)} \:. \label{cls3}
\eal
In the term \eqref{cls2} on the right-hand side, $t_{ij} (\bf{q})$
denotes the spatially transverse projector or transverse Kronecker delta
\be
t_{ij} (\bf{q}) \equiv \delta_{ij} - \hat{q}_i \hat{q}_j \:.
\ee

We shall now show, explicitly up to order $g^2$, that there is a
unique perturbative solution of the Schr\"odinger equation 
\eqref{ETSE} for the wave functional $\psi (\bf{A})$ in Eq.\
\eqref{vacans}, if we only suppose that the
dominant contribution to the coefficient function $f_k$ is at least of
order $g^{k-2}$ for $k \geq 2$. A similar method for the determination
of the vacuum wave functional has been applied before in Refs.\ 
\cite{Hat92}--\cite{MS99} to a scalar theory and to Yang-Mills theory in Weyl 
gauge.

We will consider the case
\emph{without} external charges to begin with, and include charges 
$\rho_q$ later on in the context of the static potential. To order $g^0$,
the Schr\"odinger equation reads
\be
  (N_c^2 - 1) \left( \int \frac{d^3 p}{(2 \pi)^3} f_2 (\bf{p}) \right) 
  (2 \pi)^3 \delta (\bf{0}) + \frac{1}{2} \int \frac{d^3 p}{(2 \pi)^3} 
  A_i^a (-\bf{p}) \left[ \bf{p}^2 - \big( f_2 (\bf{p}) \big)^2
  \right] A_i^a (\bf{p}) = E_0 \:, \label{g0eq}
\ee
where we have used that
\be
f_{2;ij}^{ab} (\bf{p}, -\bf{p}) = f_2 (\bf{p}) \delta_{ij} \delta^{ab}
= f_2 (-\bf{p}) \delta_{ij} \delta^{ab}
\ee
(to be contracted with spatially transverse fields) as a consequence of
the symmetry under the exchange of the arguments, of spatially rotational
and global gauge symmetry, and of the fact that $f_{2;ij}^{ab} (\bf{p}_1, 
\bf{p}_2)$ is only defined for $\bf{p}_1 + \bf{p}_2 = 0$. Equation
\eqref{g0eq} implies that, to the current order,
\bga
f_2 (\bf{p}) = \abs{\bf{p}} \:, \label{g0solf2} \\
E_0 = (N_c^2 - 1) \left( \int \frac{d^3 p}{(2 \pi)^3} \abs{\bf{p}} \right)
(2 \pi)^3 \delta (\bf{0}) \:. \label{solE0}
\ega
Generally, the energy $E_0$ cancels any field-independent terms
multiplying the vacuum functional in the Schr\"odinger equation to any
order in $g$. Eqs.\ \eqref{g0solf2} and \eqref{solE0} represent
nothing but the well-known solution of the free 
($g=0$) theory. The choice of the sign in Eq.\ 
\eqref{g0solf2} is dictated by the normalizability of the wave functional 
\eqref{vacans} to order $g^0$. As usual, $(2 \pi)^3 \delta (\bf{0})$ is 
to be understood as the total volume of space.

To the next (first) order of $g$, the Schr\"odinger equation is not much 
more complicated: it reads
\bmu
  \frac{1}{3!} \int \frac{d^3 p_1}{(2 \pi)^3}
  \frac{d^3 p_2}{(2 \pi)^3} \frac{d^3 p_3}{(2 \pi)^3} \Big\{ i g f^{abc} 
  \left[ \delta_{jk} (p_{1,l} - p_{2,l}) + \delta_{kl} (p_{2,j} - p_{3,j})
  + \delta_{lj} (p_{3,k} - p_{1,k}) \right] \\
  {}- 3 \abs{\bf{p}_1} f_{3;jkl}^{abc} (-\bf{p}_1, -\bf{p}_2, -\bf{p}_3)
  \Big\} A^a_j (\bf{p}_1) A^b_k (\bf{p}_2) A^c_l (\bf{p}_3)
  (2 \pi)^3 \delta (\bf{p}_1 + \bf{p}_2 + \bf{p}_3) = 0 \:, \label{g1eq}
\emu
where we have already taken into account the results \eqref{g0solf2},
\eqref{solE0} and the fact that
\be
f_{3;ijk}^{abc} (\bf{p}_1, \bf{p}_2, \bf{p}_3) = f^{abc} f_{3;ijk}
(\bf{p}_1, \bf{p}_2, \bf{p}_3) \:,
\ee
which is a consequence of global gauge symmetry and the invariance of the
vacuum wave functional under charge conjugation. The unique solution of
Eq.\ \eqref{g1eq} with the full symmetry under the exchange of the
arguments is
\be
  f_{3;ijk}^{abc} (\bf{p}_1, \bf{p}_2, \bf{p}_3) = -\frac{i g f^{abc}}
  {\abs{\bf{p}_1} + \abs{\bf{p}_2} + \abs{\bf{p}_3}}
  \left[ \delta_{ij} (p_{1,k} - p_{2,k}) + \delta_{jk} (p_{2,i} - p_{3,i})
  + \delta_{ki} (p_{3,j} - p_{1,j}) \right] \:. \label{solf3}
\ee
This equality, and all the following equalities with explicit spatial 
(Lorentz) indices, are proper equalities only after contracting 
with the corresponding number of transverse vector fields $\bf{A}$, or,
equivalently, after contracting every external spatial index with a transverse 
projector, for example in Eq.\ \eqref{solf3} the index $i$ with 
$t_{il} (\bf{p}_1)$.

We will now consider the Schr\"odinger equation to order
$g^2$. On the left-hand side, terms with four and two powers of $\bf{A}$
appear, which have to cancel separately, and an $\bf{A}$-independent term
which must equal $E_0$ to this order. We begin with the term with four
powers of $\bf{A}$. The quartic coupling \eqref{clA4} in the Hamiltonian
has to be cancelled by terms stemming from the second functional derivative
in Eq.\ \eqref{clA2} acting on the vacuum wave functional, and a contribution
from Eq.\ \eqref{cls3}. To order $g^2$, the coefficient functions
$f_2$ of Eq.\ \eqref{g0solf2} and $f_3$ of Eq.\ \eqref{solf3} contribute,
as well as the function $f_4$ which we will determine. As a result, the 
coefficient function $f_4$ in the vacuum wave functional takes the following 
(fully symmetric) form to order $g^2$:
\bal
\lefteqn{(|\bf{p}_1| + \ldots + |\bf{p}_4|) f_{4;ijkl}^{abcd} 
(\bf{p}_1, \ldots, \bf{p}_4)} \hspace{0.3cm} \n \\
&= g^2 \big[ 
  f^{abe} f^{cde} (\delta_{ik} \delta_{jl} - \delta_{il} \delta_{jk}) +
  f^{ace} f^{bde} (\delta_{ij} \delta_{kl} - \delta_{il} \delta_{jk}) +
  f^{ade} f^{bce} (\delta_{ij} \delta_{kl} - \delta_{ik} \delta_{jl})
  \big] \label{solf4bv} \\[1mm]
&\phantom{=} - \big[ 
f^{abe}_{3;ijm} (\bf{p}_1, \bf{p}_2, -\bf{p}_1 - \bf{p}_2) t_{mn} 
(\bf{p}_1 + \bf{p}_2) 
f^{cde}_{3;kln} (\bf{p}_3, \bf{p}_4, \bf{p}_1 + \bf{p}_2) \n \\
&\phantom{={}-} \hspace{1.5cm} {}+ 
f^{ace}_{3;ikm} (\bf{p}_1, \bf{p}_3, -\bf{p}_1 - \bf{p}_3) 
t_{mn} (\bf{p}_1 + \bf{p}_3) 
f^{bde}_{3;jln} (\bf{p}_2, \bf{p}_4, \bf{p}_1 + \bf{p}_3) \n \\
&\phantom{={}-} \hspace{3cm} {}+ 
f^{ade}_{3;ilm} (\bf{p}_1, \bf{p}_4, -\bf{p}_1 - \bf{p}_4) t_{mn} 
(\bf{p}_1 + \bf{p}_4) f^{bce}_{3;jkn} (\bf{p}_2, \bf{p}_3, \bf{p}_1 + \bf{p}_4)
\big] \label{solf4gl} \\[1mm]
&\phantom{=} - g^2 \left(
f^{abe} f^{cde} \, \delta_{ij} \delta_{kl} \, \frac{(|\bf{p}_1| - |\bf{p}_2|)
(|\bf{p}_3| - |\bf{p}_4|)}{(\bf{p}_1 + \bf{p}_2)^2} \right. \n \\
&\phantom{={}-} \hspace{2.7cm} {}+ f^{ace} f^{bde} \, \delta_{ik} \delta_{jl} 
\, \frac{(|\bf{p}_1| - |\bf{p}_3|)
(|\bf{p}_2| - |\bf{p}_4|)}{(\bf{p}_1 + \bf{p}_3)^2} \n \\
&\phantom{={}-} \hspace{5.4cm} \left. {}+ 
f^{ade} f^{bce} \, \delta_{il} \delta_{jk} \, \frac{(|\bf{p}_1| - |\bf{p}_4|)
(|\bf{p}_2| - |\bf{p}_3|)}{(\bf{p}_1 + \bf{p}_4)^2} \right) \:. \label{solf4ct}
\eal
This result for $f_4$ is represented diagrammatically in Fig.\ 
\ref{figf4}.
\begin{figure}
\begin{equation*}
2 f_4 =
- \parbox{1cm}{\begin{center}
\pspicture(-0.4,-0.4)(0.4,0.4)
\pscoil[coilaspect=35,coilwidth=0.1,coilarmA=0.1,coilarmB=0](0,0)(0.38,0.38)
\pscoil[coilaspect=35,coilwidth=0.1,coilarmA=0.1,coilarmB=0](0,0)(0.38,-0.38)
\pscoil[coilaspect=35,coilwidth=0.1,coilarmA=0.1,coilarmB=0](0,0)(-0.38,0.38)
\pscoil[coilaspect=35,coilwidth=0.1,coilarmA=0.1,coilarmB=0](0,0)(-0.38,-0.38)
\psdots(0,0)
\endpspicture
\end{center}}
- \bigg( \parbox{1.6cm}{\begin{center}
\pspicture(-0.4,-0.4)(0.97,0.4)
\pscoil[coilarmA=0.1,coilarmB=0](0.57,0)(0.95,0.38)
\pscoil[coilarmA=0.1,coilarmB=0](0.57,0)(0.95,-0.38)
\pscoil[coilarm=0.1](0,0)(0.57,0)
\pscoil[coilarmA=0.1,coilarmB=0](0,0)(-0.38,0.38)
\pscoil[coilarmA=0.1,coilarmB=0](0,0)(-0.38,-0.38)
\psdots(0,0)(0.57,0)
\endpspicture
\end{center}}
+ \text{2 perms.} \bigg)
- \bigg( \parbox{1.6cm}{\begin{center}
\pspicture(-0.4,-0.4)(0.97,0.4)
\pscoil[coilarmA=0.1,coilarmB=0](0.57,0)(0.95,0.38)
\pscoil[coilarmA=0.1,coilarmB=0](0.57,0)(0.95,-0.38)
\psline[doubleline=true,linewidth=0.8pt,doublesep=0.6pt](0,0)(0.57,0)
\pscoil[coilarmA=0.1,coilarmB=0](0,0)(-0.38,0.38)
\pscoil[coilarmA=0.1,coilarmB=0](0,0)(-0.38,-0.38)
\psdots(0,0)(0.57,0)
\endpspicture
\end{center}}
+ \text{2 perms.} \bigg)
\end{equation*}
\caption{A diagrammatic representation of Eqs.\ 
\eqref{solf4bv}--\eqref{solf4ct}. Every diagram corresponds to precisely one
of the Eqs.\ \eqref{solf4bv}--\eqref{solf4ct}, in the same order. The
``2 perms.''\ refer to permutations of the external legs. \label{figf4}}
\end{figure}
Equation \eqref{solf4bv}, divided by $(|\bf{p}_1| + \ldots + 
|\bf{p}_4|)$, is interpreted as the elementary or ``bare'' four-gluon vertex.
The r\^ole of the factor 2 and the signs in Fig.\ \ref{figf4} will become 
clear in the next section. Equation \eqref{solf4gl} and the second diagram in 
Fig.\ \ref{figf4} represent the contraction of two elementary three-gluon
vertices, the latter being given mathematically by Eq.\ \eqref{solf3}.
The contraction refers to spatial and color indices and the momenta, with 
opposite signs. Note that there is no ``propagator'' factor associated with 
the contraction (except for a transverse Kronecker delta), and there is a 
factor $1/(|\bf{p}_1| + \ldots + |\bf{p}_4|)$ for the external momenta which 
is unusual from a diagrammatic point of view. Finally, Eq.\ \eqref{solf4ct} 
and the last diagram in Fig.\ \ref{figf4} describe an
``elementary'' Coulomb interaction between the external gluon lines.

With this result in hand, we can go on to consider the terms quadratic in
$\bf{A}$ in the Schr\"odinger equation to order $g^2$. The relevant
contributions originate from Eqs.\ \eqref{clA2}, \eqref{cls1}, \eqref{cls2},
and \eqref{cls3}, and involve the functions $f_2$ and $f_4$. 
We obtain the following equation for the coefficient function $f_2$
to order $g^2$:
\bal
\lefteqn{\big( f_2 (\bf{p}) \big)^2 \delta^{ab} \delta_{ij}
= \left( \bf{p}^2 - \frac{N_c g^2}{2} \, |\bf{p}| \int \frac{d^3 q}{(2 \pi)^3} 
\frac{1 - (\hat{\bf{p}} \cdot \hat{\bf{q}})^2}{(\bf{p} - \bf{q})^2} \right) 
\delta^{ab} \delta_{ij}} \hspace{1.5cm} \n \\
&{}+ \frac{1}{2} \int \frac{d^3 q}{(2 \pi)^3} \, 
f_{4;ijkl}^{abcc} (-\bf{p}, \bf{p}, -\bf{q}, \bf{q}) 
t_{kl} (\bf{q})
- N_c g^2 \delta^{ab} \int \frac{d^3 q}{(2 \pi)^3} \frac{|\bf{p}| - |\bf{q}|}
{(\bf{p} - \bf{q})^2} \, t_{ij} (\bf{q}) \:. \label{g2solf2gen}
\eal
The explicit expression for $f_2$ to order $g^2$ is
\bal
f_2 (\bf{p}) &= |\bf{p}| - \frac{N_c g^2}{4} \int \frac{d^3 q}{(2 \pi)^3} 
\frac{1 - (\hat{\bf{p}} \cdot \hat{\bf{q}})^2}{(\bf{p} - \bf{q})^2} 
+ \frac{N_c g^2}{2 |\bf{p}|} \, \frac{4}{3} \int \frac{d^3 q}{(2 \pi)^3} 
\frac{1}{2 |\bf{p}| + 2 |\bf{q}|} \label{g2solf2tp} \\
&\phantom{=} {}- \frac{N_c g^2}{2 |\bf{p}|} \, 2 \int \frac{d^3 q}{(2 \pi)^3}
\frac{\big( \delta_{ik} p_l + \delta_{kl} q_i - \delta_{li} p_k \big)
\, t_{km} (\bf{p} - \bf{q}) \, t_{ln} (\bf{q})}
{2 |\bf{p}| + 2 |\bf{q}|} \n \\
&\hspace{3.5cm} {}\times \frac{\big( \delta_{jm} p_n + \delta_{mn} q_j 
- \delta_{nj} p_m \big) \, t_{ij} (\bf{p})}
{(|\bf{p}| + |\bf{q}| + |\bf{p} - \bf{q}|)^2} \label{g2solf2gl} \\
&\phantom{=} {}- \frac{N_c g^2}{2 |\bf{p}|} \, \frac{1}{2} 
\int \frac{d^3 q}{(2 \pi)^3} \frac{1 + (\hat{\bf{p}} \cdot \hat{\bf{q}})^2}
{2 |\bf{p}| + 2 |\bf{q}|} \, \frac{\left( |\bf{p}| - |\bf{q}| \right)^2}
{(\bf{p} - \bf{q})^2} \label{g2solf2ctl} \\
&\phantom{=} {}- \frac{N_c g^2}{2 |\bf{p}|} \, \frac{1}{2} 
\int \frac{d^3 q}{(2 \pi)^3}
\left( 1 + (\hat{\bf{p}} \cdot \hat{\bf{q}})^2 \right)
\frac{|\bf{p}| - |\bf{q}|}{(\bf{p} - \bf{q})^2} \:, \label{g2solf2ct}
\eal
where we have used the contraction of an arbitrary tensor $T_{ij} (\bf{p})$
\be
\frac{1}{2} \, T_{ij} (\bf{p}) t_{ij} (\bf{p}) = T^t (\bf{p})
\ee
in order to extract the transverse part. We have presented the 
diagrams corresponding to Eqs.\ \eqref{g2solf2tp}--\eqref{g2solf2ct} in 
Fig.\ \ref{figf2}.
\begin{figure}
\begin{equation*}
2 f_2 =
\big( 
\parbox{0.74cm}{\begin{center}
\pspicture(0,-0.25)(0.74,0.25)
\psCoil{50}{3190}
\psdots[dotstyle=o,dotscale=1.1](0.37,0)
\endpspicture
\end{center}}
\big)^{-1}
%
%
- \parbox{2.1cm}{\begin{center}
\pspicture(-0.6,-0.5)(1.5,0.5)
\pscoil[coilarmA=0.1,coilarmB=0](0,0)(-0.47,0)
\pscoil[coilarmA=0.1,coilarmB=0](0.9,0)(1.37,0)
\pscircle[linewidth=0.5pt,linestyle=dashed,dash=2pt 1.5pt](0.45,0){0.45}
\psdots(0,0)(0.9,0)
\endpspicture
\end{center}}
- \parbox{1.7cm}{\begin{center}
\pspicture(-0.85,-0.55)(0.85,0.5)
\SpecialCoor
\multido{\n=-73.00+11.25}{31}{%
  \FPadd{-90}{\n}{\m}
  \rput{\m}(0.4;\n){\psCoil{-65}{425}}}
\pscoil[coilarmA=0.1,coilarmB=0](0,-0.45)(-0.715,-0.45)
\pscoil[coilarmA=0.1,coilarmB=0](0,-0.45)(0.715,-0.45)
\psdots(0,-0.45)
\endpspicture
\end{center}}
- \parbox{2.2cm}{\begin{center}
\pspicture(-1.05,-0.5)(1.05,0.5)
\SpecialCoor
\multido{\n=17.00+11.25}{15}{%
  \FPadd{-90}{\n}{\m}
  \rput{\m}(0.4;\n){\psCoil{-65}{425}}}
\multido{\n=197.00+11.25}{15}{%
  \FPadd{-90}{\n}{\m}
  \rput{\m}(0.4;\n){\psCoil{-65}{425}}}
\pscoil[coilarmA=0.1,coilarmB=0](-0.45,0)(-0.92,0)
\pscoil[coilarmA=0.1,coilarmB=0](0.45,0)(0.92,0)
\psdots(-0.45,0)(0.45,0)
\endpspicture
\end{center}}
- \parbox{2.2cm}{\begin{center}
\pspicture(-1.05,-0.5)(1.05,0.5)
\SpecialCoor
\multido{\n=17.00+11.25}{15}{%
  \FPadd{-90}{\n}{\m}
  \rput{\m}(0.4;\n){\psCoil{-65}{425}}}
\psline[doubleline=true,linewidth=0.8pt,doublesep=0.6pt](-0.45,0)(0.45,0)
\pscoil[coilarmA=0.1,coilarmB=0](-0.45,0)(-0.92,0)
\pscoil[coilarmA=0.1,coilarmB=0](0.45,0)(0.92,0)
\psdots(-0.45,0)(0.45,0)
\endpspicture
\end{center}}
- \parbox{2.2cm}{\begin{center}
\pspicture(-1.05,-0.5)(1.05,0.5)
\SpecialCoor
\multido{\n=17.00+11.25}{15}{%
  \FPadd{-90}{\n}{\m}
  \rput{\m}(0.4;\n){\psCoil{-65}{425}}}
\psline[doubleline=true,linewidth=0.8pt,doublesep=0.6pt](-0.45,0)(0.45,0)
\pscoil[coilarmA=0.1,coilarmB=0](-0.45,0)(-0.92,0)
\pscoil[coilarmA=0.1,coilarmB=0](0.45,0)(0.92,0)
\psdots(-0.45,0)(0.45,0)
\psdots[dotstyle=square,dotscale=1.4](0,0.4)
\psdots[dotstyle=+,dotscale=1.4,dotangle=45](0,0.4)
\endpspicture
\end{center}}
%
%
%
%
\end{equation*}
\caption{The diagrams corresponding to Eqs.\ 
\eqref{g2solf2tp}--\eqref{g2solf2ct}. The bare propagator, the inverse 
$2 |\bf{p}|$ of which appears in Eq.\ \eqref{g2solf2tp}, is marked with an
open circle for later use. The first two one-loop diagrams correspond to
the integrals in Eq.\ \eqref{g2solf2tp}, in the same order. The following
diagrams represent Eqs.\ \eqref{g2solf2gl}--\eqref{g2solf2ct}, respectively. 
See the text for a motivation of the ``crossed'' gluon propagator notation
in the last loop diagram. \label{figf2}}
\end{figure}
The first loop integral in Eqs.\ \eqref{g2solf2gen} and \eqref{g2solf2tp} 
results from Eq.\ \eqref{cls1} and is represented in Fig.\ \ref{figf2} as a 
ghost loop because it stems from the Faddeev-Popov determinant. The 
following three loop diagrams in Fig.\ \ref{figf2} are obtained by contracting
two external legs in the diagrams of Fig.\ \ref{figf4}, see Eq.\ 
\eqref{g2solf2gen}. The last loop integral in Eq.\ \eqref{g2solf2gen}, or 
the integral \eqref{g2solf2ct}, on the other hand, originates from the terms 
\eqref{cls2} and \eqref{cls3} in the Hamiltonian. Lacking a better
notation, we distinguish this contribution from the contraction of 
Eq.\ \eqref{solf4ct}, the previous diagram, by marking the gluon propagator 
with a cross (because there is no term $|\bf{q}|$ in the denominator that 
would indicate the presence of an internal gluon propagator --- in
fact, the diagram may be interpreted to contain a $\Pi \Pi$-correlator,
where $\bm{\Pi}$ is the momentum conjugate to $\bf{A}$).

We have thus completed the determination of the (exponent of the) perturbative 
vacuum wave functional to order $g^2$. The result is given in Eqs.\ 
\eqref{solf3}, \eqref{solf4bv}--\eqref{solf4ct}, and 
\eqref{g2solf2tp}--\eqref{g2solf2ct}, to be substituted in Eq.\ 
\eqref{vacans}. We can also extract the perturbative vacuum energy to the
same order from the $\bf{A}$-independent terms in the Schr\"odinger equation
with the result
\be
E_0 = (N_c^2 - 1) \left( \int \frac{d^3 p}{(2 \pi)^3} f_2 (\bf{p}) \right)
(2 \pi)^3 \delta (\bf{0}) \label{g2E0}
\ee
[cf.\ Eq.\ \eqref{g0eq}], where Eqs.\ \eqref{g2solf2tp}--\eqref{g2solf2ct} 
have to be substituted for $f_2 (\bf{p})$.
The explicit expression is not relevant for our purposes. We shall come back
to the vacuum energy later in the context of the static potential in the
presence of external charges. It should also be clear by now how to take the 
determination of the perturbative vacuum functional and the vacuum energy 
systematically to higher orders.

\section{Equal-time two-point correlation functions}

For the calculation of the equal-time correlation functions,
we need to include the Faddeev-Popov determinant in the measure of the
functional integral, see Eq.\ \eqref{corrfunc}. For our diagrammatic
procedure, it is very convenient to introduce ghost fields and write
\be
\textsf{FP} (\bf{A}) = \int D [c, \bar{c}] \exp \left( - \int d^3 x \,
\bar{c}^a (\bf{x}) [ - \nabla \cdot \bf{D}^{ab} (\bf{A}) ] c^b (\bf{x}) 
\right) \:. \label{FPrep}
\ee
In our conventions, we have explicitly
\bal
\lefteqn{\int d^3 x \, \bar{c}^a (\bf{x}) [ - \nabla \cdot 
\bf{D}^{ab} (\bf{A}) ] c^b (\bf{x}) 
= \int \frac{d^3 p}{(2 \pi)^3} \, \bar{c}^a (-\bf{p}) \, \bf{p}^2 
c^a (\bf{p})} \hspace{1cm} \label{gg} \\
&{}+ g \int \frac{d^3 p_1}{(2 \pi)^3} \frac{d^3 p_2}{(2 \pi)^3} 
\frac{d^3 p_3}{(2 \pi)^3} \, i f^{abc} \, p_{1,j} \, \bar{c}^a (\bf{p}_1) 
c^b (\bf{p}_2) A^c_j (\bf{p}_3) 
(2 \pi)^3 \delta(\bf{p}_1 + \bf{p}_2 + \bf{p}_3) \:. \label{ggv}
\eal
Note that $p_{1,j}$ under the integral in Eq.\ \eqref{ggv} can be replaced by 
$-p_{2,j}$ due to the transversality of $\bf{A}$.

We now have a representation of the equal-time correlation functions as a
functional integral over the transverse components of $\bf{A}$, the ghost
and the antighost fields, see Eq.\ \eqref{corrfunc}. The integration measure,
which would be the exponential of the negative of the Euclidean action in the 
usual four-dimensional formulation (in Euclidean space), is now given by
the exponential in Eq.\ \eqref{FPrep} and the square of Eq.\ \eqref{vacans}.
Note that the vacuum functional is real (at least to order $g^2$) because the
coefficient functions fulfill the reality condition
\be
\left( f^{a_1 \ldots a_k}_{k; i_1 \ldots i_k} (-\bf{p}_1, \ldots, -\bf{p}_k)
\right)^\ast = 
f^{a_1 \ldots a_k}_{k; i_1 \ldots i_k} (\bf{p}_1, \ldots, \bf{p}_k) \:.
\ee

We shall use the analogy of this representation with the familiar functional 
integral representation of the covariant correlation functions in the usual
four-dimensional formulation for the perturbative determination of the 
equal-time correlation functions \eqref{corrfunc}. The corresponding
Feynman rules are easily identified: the (static) gluon propagator is the 
inverse of $2 |\bf{p}|$, cf.\ Eq.\ \eqref{g0solf2} (the factor of two is due 
to the square of the wave functional in the measure), the other contributions
$-2 \big( f_2 (\bf{p}) - |\bf{p}| \big)$ and the other coefficient functions 
$-2 f_3 (\bf{p}_1, \bf{p}_2, \bf{p}_3)$ and 
$-2 f_4 (\bf{p}_1, \ldots, \bf{p}_4)$ determine the two-, three-, and 
four-gluon vertices. Furthermore, from Eq.\ \eqref{gg} we identify the free 
ghost propagator $1/ \bf{p}^2$ and from Eq.\ \eqref{ggv} the ghost-gluon 
vertex. For the case of $\phi^4$ theory in $(1+1)$ dimensions, the
calculation of equal-time correlation functions from a representation
analogous to \eqref{corrfunc} has been discussed in Ref.\ \cite{JM00}.

We consider the gluon equal-time two-point function 
$\langle A_i^a (\bf{p}_1) A_j^b (\bf{p}_2) \rangle$ 
(with $t=0$ in the arguments of the gluon fields to be understood) 
first. One of the contributions to be taken into account is the
ghost loop, constructed from two ghost-gluon vertices \eqref{ggv} and two
ghost propagators [see Eq.\ \eqref{gg}], and furthermore two static gluon 
propagators from Eq.\ \eqref{g0solf2} for the external lines. As it turns 
out, this contribution is exactly cancelled by the other contribution 
with the same graph ``topology'' which arises from contracting one of the 
two-gluon vertices, (minus twice) the first integral in Eq.\ 
\eqref{g2solf2tp}, with two external gluon propagators. Both
contributions are represented diagrammatically in the first line of Fig.\
\ref{figAA}.
\begin{figure}
\begin{align*}
\parbox{2.4cm}{\begin{center}
\pspicture(-0.75,-0.5)(1.65,0.5)
\pscoil[coilarmA=0.1,coilarmB=0](0,0)(-0.63,0)
\pscoil[coilarmA=0.1,coilarmB=0](0.9,0)(1.53,0)
\pscircle[linewidth=0.5pt,linestyle=dashed](0.45,0){0.45}
\psdots(0,0)(0.9,0)
\psdots[dotstyle=o,dotscale=1.1](-0.37,0)(1.27,0)(0.45,0.45)(0.45,-0.45)
\endpspicture
\end{center}}
+ \parbox{2.4cm}{\begin{center}
\pspicture(-0.75,-0.5)(1.65,0.5)
\pscoil[coilarmA=0.1,coilarmB=0](0,0)(-0.63,0)
\pscoil[coilarmA=0.1,coilarmB=0](0.9,0)(1.53,0)
\pscircle[linewidth=0.5pt,linestyle=dashed,dash=2pt 1.5pt](0.45,0){0.45}
\psdots(0,0)(0.9,0)
\psdots[dotstyle=o,dotscale=1.1](-0.37,0)(1.27,0)
\endpspicture
\end{center}}
&= 0 \\[-8mm]
\parbox{1.7cm}{\begin{center}
\pspicture(-0.85,-0.55)(0.85,0.5)
\SpecialCoor
\multido{\n=-73.00+11.25}{31}{%
  \FPadd{-90}{\n}{\m}
  \rput{\m}(0.4;\n){\psCoil{-65}{425}}}
\pscoil[coilarmA=0.1,coilarmB=0](0,-0.45)(-0.715,-0.45)
\pscoil[coilarmA=0.1,coilarmB=0](0,-0.45)(0.715,-0.45)
\psdots(0,-0.45)
\psdots[dotstyle=o,dotscale=1.1](-0.45,-0.45)(0.45,-0.45)(0,0.4)
\endpspicture
\end{center}}
+ \parbox{1.7cm}{\begin{center}
\pspicture(-0.85,-0.55)(0.85,0.5)
\SpecialCoor
\multido{\n=-73.00+11.25}{31}{%
  \FPadd{-90}{\n}{\m}
  \rput{\m}(0.4;\n){\psCoil{-65}{425}}}
\pscoil[coilarmA=0.1,coilarmB=0](0,-0.45)(-0.715,-0.45)
\pscoil[coilarmA=0.1,coilarmB=0](0,-0.45)(0.715,-0.45)
\psdots(0,-0.45)
\psdots[dotstyle=o,dotscale=1.1](-0.45,-0.45)(0.45,-0.45)
\endpspicture
\end{center}}
&= E \bigg( \parbox{1.5cm}{\begin{center}
\pspicture(-0.75,-0.55)(0.75,0.5)
\SpecialCoor
\multido{\n=-73.00+11.25}{31}{%
  \FPadd{-90}{\n}{\m}
  \rput{\m}(0.4;\n){\psCoil{-65}{425}}}
\pscoil[coilarmA=0.1,coilarmB=0](0,-0.45)(-0.715,-0.45)
\pscoil[coilarmA=0.1,coilarmB=0](0,-0.45)(0.715,-0.45)
\psdots(0,-0.45)
\psdots[dotstyle=o,dotscale=1.1](-0.45,-0.45)(0.45,-0.45)(0,0.4)
\endpspicture
\end{center}} \bigg)  \\[-6mm]
\parbox{2.4cm}{\begin{center}
\pspicture(-1.2,-0.5)(1.2,0.5)
\SpecialCoor
\multido{\n=17.00+11.25}{15}{%
  \FPadd{-90}{\n}{\m}
  \rput{\m}(0.4;\n){\psCoil{-65}{425}}}
\multido{\n=197.00+11.25}{15}{%
  \FPadd{-90}{\n}{\m}
  \rput{\m}(0.4;\n){\psCoil{-65}{425}}}
\pscoil[coilarmA=0.1,coilarmB=0](-0.45,0)(-1.08,0)
\pscoil[coilarmA=0.1,coilarmB=0](0.45,0)(1.08,0)
\psdots(-0.45,0)(0.45,0)
\psdots[dotstyle=o,dotscale=1.1](-0.82,0)(0.82,0)(0,0.4)(0,-0.4)
\endpspicture
\end{center}}
+ \parbox{2.4cm}{\begin{center}
\pspicture(-1.2,-0.5)(1.2,0.5)
\SpecialCoor
\multido{\n=17.00+11.25}{15}{%
  \FPadd{-90}{\n}{\m}
  \rput{\m}(0.4;\n){\psCoil{-65}{425}}}
\multido{\n=197.00+11.25}{15}{%
  \FPadd{-90}{\n}{\m}
  \rput{\m}(0.4;\n){\psCoil{-65}{425}}}
\pscoil[coilarmA=0.1,coilarmB=0](-0.45,0)(-1.08,0)
\pscoil[coilarmA=0.1,coilarmB=0](0.45,0)(1.08,0)
\psdots(-0.45,0)(0.45,0)
\psdots[dotstyle=o,dotscale=1.1](-0.82,0)(0.82,0)(0,0.4)
\endpspicture
\end{center}}
+ \parbox{2.4cm}{\begin{center}
\pspicture(-1.2,-0.5)(1.2,0.5)
\SpecialCoor
\multido{\n=17.00+11.25}{15}{%
  \FPadd{-90}{\n}{\m}
  \rput{\m}(0.4;\n){\psCoil{-65}{425}}}
\multido{\n=197.00+11.25}{15}{%
  \FPadd{-90}{\n}{\m}
  \rput{\m}(0.4;\n){\psCoil{-65}{425}}}
\pscoil[coilarmA=0.1,coilarmB=0](-0.45,0)(-1.08,0)
\pscoil[coilarmA=0.1,coilarmB=0](0.45,0)(1.08,0)
\psdots(-0.45,0)(0.45,0)
\psdots[dotstyle=o,dotscale=1.1](-0.82,0)(0.82,0)
\endpspicture
\end{center}}
&= E \bigg( \parbox{2.24cm}{\begin{center}
\pspicture(-1.12,-0.5)(1.12,0.5)
\SpecialCoor
\multido{\n=17.00+11.25}{15}{%
  \FPadd{-90}{\n}{\m}
  \rput{\m}(0.4;\n){\psCoil{-65}{425}}}
\multido{\n=197.00+11.25}{15}{%
  \FPadd{-90}{\n}{\m}
  \rput{\m}(0.4;\n){\psCoil{-65}{425}}}
\pscoil[coilarmA=0.1,coilarmB=0](-0.45,0)(-1.08,0)
\pscoil[coilarmA=0.1,coilarmB=0](0.45,0)(1.08,0)
\psdots(-0.45,0)(0.45,0)
\psdots[dotstyle=o,dotscale=1.1](-0.82,0)(0.82,0)(0,0.4)(0,-0.4)
\endpspicture
\end{center}} \bigg) \\[-6mm]
\parbox{2.4cm}{\begin{center}
\pspicture(-1.2,-0.5)(1.2,0.5)
\SpecialCoor
\multido{\n=17.00+11.25}{15}{%
  \FPadd{-90}{\n}{\m}
  \rput{\m}(0.4;\n){\psCoil{-65}{425}}}
\psline[doubleline=true,linewidth=0.8pt,doublesep=0.6pt](-0.45,0)(0.45,0)
\pscoil[coilarmA=0.1,coilarmB=0](-0.45,0)(-1.08,0)
\pscoil[coilarmA=0.1,coilarmB=0](0.45,0)(1.08,0)
\psdots(-0.45,0)(0.45,0)
\psdots[dotstyle=o,dotscale=1.1](-0.82,0)(0.82,0)(0,0.4)
\endpspicture
\end{center}}
+ \parbox{2.4cm}{\begin{center}
\pspicture(-1.2,-0.5)(1.2,0.5)
\SpecialCoor
\multido{\n=17.00+11.25}{15}{%
  \FPadd{-90}{\n}{\m}
  \rput{\m}(0.4;\n){\psCoil{-65}{425}}}
\psline[doubleline=true,linewidth=0.8pt,doublesep=0.6pt](-0.45,0)(0.45,0)
\pscoil[coilarmA=0.1,coilarmB=0](-0.45,0)(-1.08,0)
\pscoil[coilarmA=0.1,coilarmB=0](0.45,0)(1.08,0)
\psdots(-0.45,0)(0.45,0)
\psdots[dotstyle=o,dotscale=1.1](-0.82,0)(0.82,0)
\endpspicture
\end{center}}
+ \parbox{2.4cm}{\begin{center}
\pspicture(-1.2,-0.5)(1.2,0.5)
\SpecialCoor
\multido{\n=17.00+11.25}{15}{%
  \FPadd{-90}{\n}{\m}
  \rput{\m}(0.4;\n){\psCoil{-65}{425}}}
\psline[doubleline=true,linewidth=0.8pt,doublesep=0.6pt](-0.45,0)(0.45,0)
\pscoil[coilarmA=0.1,coilarmB=0](-0.45,0)(-1.08,0)
\pscoil[coilarmA=0.1,coilarmB=0](0.45,0)(1.08,0)
\psdots(-0.45,0)(0.45,0)
\psdots[dotstyle=o,dotscale=1.1](-0.82,0)(0.82,0)
\psdots[dotstyle=square,dotscale=1.4](0,0.4)
\psdots[dotstyle=+,dotscale=1.4,dotangle=45](0,0.4)
\endpspicture
\end{center}}
&= E \bigg( \parbox{2.24cm}{\begin{center}
\pspicture(-1.12,-0.5)(1.12,0.5)
\SpecialCoor
\multido{\n=17.00+11.25}{15}{%
  \FPadd{-90}{\n}{\m}
  \rput{\m}(0.4;\n){\psCoil{-65}{425}}}
\psline[doubleline=true,linewidth=0.8pt,doublesep=0.6pt](-0.45,0)(0.45,0)
\pscoil[coilarmA=0.1,coilarmB=0](-0.45,0)(-1.08,0)
\pscoil[coilarmA=0.1,coilarmB=0](0.45,0)(1.08,0)
\psdots(-0.45,0)(0.45,0)
\psdots[dotstyle=o,dotscale=1.1](-0.82,0)(0.82,0)(0,0.4)
\endpspicture
\end{center}} \bigg)
+ \parbox{2.4cm}{\begin{center}
\pspicture(-1.2,-0.5)(1.2,0.5)
\SpecialCoor
\multido{\n=17.00+11.25}{15}{%
  \FPadd{-90}{\n}{\m}
  \rput{\m}(0.4;\n){\psCoil{-65}{425}}}
\psline[doubleline=true,linewidth=0.8pt,doublesep=0.6pt](-0.45,0)(0.45,0)
\pscoil[coilarmA=0.1,coilarmB=0](-0.45,0)(-1.08,0)
\pscoil[coilarmA=0.1,coilarmB=0](0.45,0)(1.08,0)
\psdots(-0.45,0)(0.45,0)
\psdots[dotstyle=o,dotscale=1.1](-0.82,0)(0.82,0)
\psdots[dotstyle=square,dotscale=1.4](0,0.4)
\psdots[dotstyle=+,dotscale=1.4,dotangle=45](0,0.4)
\endpspicture
\end{center}}
\end{align*}
\caption{Diagrammatic representation of the various contributions to the
gluonic equal-time two-point function, see Eqs.\ 
\eqref{exprepdet}--\eqref{AActsum}. The propagators marked with open circles 
are taken from Eqs.\ \eqref{g0solf2} and \eqref{gg}, respectively, while the
``direct'' contractions without open circles refer to the contractions
that appear in the course of the determination of the vacuum wave functional,
see Figs.\ \ref{figf4} and \ref{figf2}, so that the corresponding parts of
the diagrams translate into (minus two times) the mathematical expressions
\eqref{solf4gl}--\eqref{solf4ct} [divided by $(|\bf{p}_1| + \ldots + 
|\bf{p}_4|)$] and \eqref{g2solf2tp}--\eqref{g2solf2ct}. The notation
``$E (\cdot)$'' for the sum of all diagrams with the same topology is
explained in the text, following Eq.\ \eqref{g2AAct}. \label{figAA}}
\end{figure}
The cancellation of the ghost loop contribution from the
perturbative gluon two-point function (to one-loop order) is interesting
given that this contribution plays a major role in the nonperturbative
approaches to the infrared behavior of the equal-time gluon two-point 
function \cite{Zwa04}--\cite{SLR06}. The cancellation of ghost loops was found 
to be a general feature in the Lagrangian functional integral approach 
\cite{WR07a,WR07b}. An alternative way to see the cancellation in our present 
approach is to write the Faddeev-Popov determinant as
\be
\textsf{FP} (\bf{A}) = \exp \big[ \text{tr} \, \ln \big( -\nabla \cdot 
\bf{D} (\bf{A}) \big) \big] \:. \label{exprepdet}
\ee
The coefficient of the term quadratic in $\bf{A}$ in $\text{tr} \, \ln 
\big( -\nabla \cdot \bf{D} (\bf{A}) \big)$ precisely equals twice the first 
integral in Eq.\ \eqref{g2solf2tp} and hence cancels out in the exponent.

Next we turn to the tadpole contribution which is obtained from the elementary
four-gluon vertex extracted from Eq.\ \eqref{solf4bv} appropriately contracted
with three static gluon propagators. Again, there is a second contribution
with the same ``topology'' given by the two-gluon vertex from the last integral
in Eq.\ \eqref{g2solf2tp} contracted with two external propagators, 
cf.\ the second line in Fig.\ \ref{figAA}. The sum of these two 
contributions to the gluon equal-time two-point function is (with 
$\bf{p} \equiv \bf{p}_1$)
\bmu
-\frac{2 N_c g^2}{(2 |\bf{p}|)^2} \, \frac{4}{3} \int \frac{d^3 q}{(2 \pi)^3}
\frac{1}{2 |\bf{p}| + 2 |\bf{q}|} \frac{1}{2 |\bf{q}|}
- \frac{2 N_c g^2}{(2 |\bf{p}|)^3} \, \frac{4}{3} \int \frac{d^3 q}{(2 \pi)^3}
\frac{1}{2 |\bf{p}| + 2 |\bf{q}|} \\
= -\frac{2 N_c g^2}{(2 |\bf{p}|)^3} \, \frac{4}{3} \int \frac{d^3 q}{(2 \pi)^3}
\frac{1}{2 |\bf{q}|} \:, \label{AAtpsum}
\emu
to be multiplied with $\delta^{ab} t_{ij} (\bf{p})$.

The most complicated contribution to the two-point function comes from diagrams
with the gluon loop topology (with two three-gluon vertices). There are three
different diagrams of this type, represented in the third line of
Fig.\ \ref{figAA}, the first from contracting two three-gluon 
vertices extracted from Eq.\ \eqref{solf3} with four static gluon propagators
(two internal and two external), the second from contracting the part of
the four-gluon vertex given by Eq.\ \eqref{solf4gl} with one internal and two
external gluon propagators, and the third by contracting the two-gluon vertex 
from Eq.\ \eqref{g2solf2gl} with two static gluon propagators. The sum of 
these contributions is
\bmu
\frac{2 N_c g^2}{(2 |\bf{p}|)^3} \, 2 \int \frac{d^3 q}{(2 \pi)^3}
\frac{2 |\bf{p}| + 2 |\bf{p} - \bf{q}|}
{(|\bf{p}| + |\bf{q}| + |\bf{p} - \bf{q}|)^2 \, 2 |\bf{q}| \, 
2 |\bf{p} - \bf{q}|} \\
{}\times \big( \delta_{km} p_n + \delta_{mn} q_k - \delta_{nk} p_m \big) \,
t_{mr} (\bf{p} - \bf{q}) \, t_{ns} (\bf{q}) \,
\big( \delta_{lr} p_s + \delta_{rs} q_l 
- \delta_{sl} p_r \big) \, t_{kl} (\bf{p}) \label{AAglsum}
\emu
(again, to be multiplied with $\delta^{ab} t_{ij} (\bf{p})$,
and $\bf{p} \equiv \bf{p}_1$). The tensor structure in this expression is
invariant under the transformation $\bf{q} \to \bf{p} - \bf{q}$, a
fact we can use to replace $2 |\bf{p}| + 2 |\bf{p} - \bf{q}|$ in the 
numerator with $2 |\bf{p}| + |\bf{q}| + |\bf{p} - \bf{q}|$. The tensor 
structure itself can be simplified by performing the contractions explicitly. 
A straightforward, but somewhat tedious calculation gives
\bmu
\big( \delta_{km} p_n + \delta_{mn} q_k - \delta_{nk} p_m \big) \,
t_{mr} (\bf{p} - \bf{q}) \, t_{ns} (\bf{q}) \,
\big( \delta_{lr} p_s + \delta_{rs} q_l 
- \delta_{sl} p_r \big) \, t_{kl} (\bf{p}) \\
= \Big( 1 - (\hat{\bf{p}} \cdot \hat{\bf{q}})^2 \Big)
\left( 2 \bf{p}^2 + 2 \bf{q}^2 + \frac{\bf{p}^2 \, \bf{q}^2 + 
(\bf{p} \cdot \bf{q})^2}{(\bf{p} - \bf{q})^2} \right) \:.
\emu

Finally, we turn to the contributions that involve the (non-abelian) Coulomb 
potential. There are, again, three such terms, represented in the last 
line of Fig.\ \ref{figAA}, the first from the four-gluon vertex
derived from Eq.\ \eqref{solf4ct} contracted with three static gluon
propagators, and the other two using the two-gluon vertices
corresponding to the two integrals \eqref{g2solf2ctl} and 
\eqref{g2solf2ct} contracted with two gluon propagators each. The
sum of these terms is
\bmu
\frac{2 N_c g^2}{(2 |\bf{p}|)^3} \, \frac{1}{2} 
\int \frac{d^3 q}{(2 \pi)^3} \frac{1 + (\hat{\bf{p}} \cdot \hat{\bf{q}})^2}
{2 |\bf{q}|} \, \frac{(|\bf{p}| - |\bf{q}|)^2}{(\bf{p} - \bf{q})^2}
+ \frac{2 N_c g^2}{(2 |\bf{p}|)^3} \, \frac{1}{2} 
\int \frac{d^3 q}{(2 \pi)^3} \left( 1 + (\hat{\bf{p}} \cdot \hat{\bf{q}})^2
\right) \frac{|\bf{p}| - |\bf{q}|}{(\bf{p} - \bf{q})^2} \\
= \frac{2 N_c g^2}{(2 |\bf{p}|)^3} \, \frac{1}{2} 
\int \frac{d^3 q}{(2 \pi)^3} \frac{1 + (\hat{\bf{p}} \cdot \hat{\bf{q}})^2}
{2 |\bf{q}|} \, \frac{|\bf{p}|^2 - |\bf{q}|^2}{(\bf{p} - \bf{q})^2} \:,
\label{AActsum}
\emu
to be multiplied with $\delta^{ab} t_{ij} (\bf{p})$ as before,
and $\bf{p} \equiv \bf{p}_1$. On the left-hand side of Eq.\ 
\eqref{AActsum}, we have added up the contributions from the contraction
of the four-gluon vertex and the two-gluon vertex in Eq.\ \eqref{g2solf2ctl}
to give the first loop integral. The left-hand side of Eq.\ \eqref{AActsum} 
is represented diagrammatically as the right-hand side of the last line in
Fig.\ \ref{figAA}.

Putting it all together, the result for the equal-time gluon two-point function
is, to order $g^2$,
\bal
\lefteqn{\langle A_i^a (\bf{p}_1) A_j^b (\bf{p}_2) \rangle 
= \left[ \frac{1}{2 |\bf{p}_1|} - \frac{2 N_c g^2}{(2 |\bf{p}_1|)^3} \, 
\frac{4}{3} \int \frac{d^3 q}{(2 \pi)^3} \frac{1}{2 |\bf{q}|} \right.} 
\hspace{1.5cm} \label{g2AAtp} \\
&\phantom{=\:\;} {}+ \frac{2 N_c g^2}{(2 |\bf{p}_1|)^3} \, 2 
\int \frac{d^3 q}{(2 \pi)^3}
\frac{\big( 1 - (\hat{\bf{p}}_1 \cdot \hat{\bf{q}})^2 \big)
(2 |\bf{p}_1| + |\bf{q}| + |\bf{p}_1 - \bf{q}|)}
{(|\bf{p}_1| + |\bf{q}| + |\bf{p}_1 - \bf{q}|)^2 \, 2 |\bf{q}| \, 
2 |\bf{p}_1 - \bf{q}|} \n \\
&\phantom{=} \hspace{4cm} {}\times
\left( 2 \bf{p}_1^2 + 2 \bf{q}^2 + \frac{\bf{p}_1^2 \, \bf{q}^2 + 
(\bf{p}_1 \cdot \bf{q})^2}{(\bf{p}_1 - \bf{q})^2} \right) \label{g2AAgl} \\
&\phantom{=\:} \left. {}+ \frac{2 N_c g^2}{(2 |\bf{p}_1|)^3} \, \frac{1}{2} 
\int \frac{d^3 q}{(2 \pi)^3} \frac{1 + (\hat{\bf{p}}_1 \cdot \hat{\bf{q}})^2}
{2 |\bf{q}|} \, \frac{\bf{p}_1^2 - \bf{q}^2}{(\bf{p}_1 - \bf{q})^2} \right]
\delta^{ab} \, t_{ij} (\bf{p}_1) 
(2 \pi)^3 \delta (\bf{p}_1 + \bf{p}_2) \:. \label{g2AAct}
\eal
Looking at the diagrams on the left-hand sides of the first three 
lines in Fig.\ \ref{figAA}, it is clear that there is precisely one diagram 
among those of the same topology that is contructed exclusively from the 
elementary vertices \eqref{solf3}, \eqref{solf4bv}, and \eqref{ggv} and the 
``bare'' propagators taken from Eqs.\ \eqref{g0solf2} and \eqref{gg}. We will 
call this kind of diagram an ``F-diagram''. Interestingly, the sum of all
diagrams with the same topology which we will refer to as an ``E-diagram'',
can be constructed from the corresponding F-diagram by a formal operation
that we call the ``E-operator''. It is denoted as ``$E (\cdot)$'' in Fig.\
\ref{figAA} and will be illustrated considering the example of the third
line in Fig.\ \ref{figAA}: starting from the mathematical expression for
the corresponding F-diagram, one multiplies the integrand (of the integral over
loop momentum) with the sum of all $|\bf{k}|$ where $\bf{k}$ runs over the
momenta of all propagators in the diagram, and divides by the 
corresponding sum restricted to the momenta of the external propagators. 
Indeed, the result of this operation is given by Eq.\ \eqref{AAglsum} in the 
$(\bf{q} \to \bf{p} - \bf{q})$-symmetric form, see the remark after Eq.\ 
\eqref{AAglsum}.

The same rule applies to the second line in Fig.\ \ref{figAA},
or Eq.\ \eqref{AAtpsum}, only that the propagator that starts and ends at
the same vertex has to be counted twice in the sum over (internal and
external) $|\bf{k}|$. Similarly, the E-operator can be used to sum the
first two diagrams on the left-hand side of the last line in Fig.\ 
\ref{figAA}. We have to consider the Coulomb interaction as an elementary
vertex given by \eqref{solf4ct} to this end, and count the gluon propagator 
that starts and ends at this vertex twice in the sum over $|\bf{k}|$ just as
in the case of the other elementary four-gluon vertex. The same rule for the
generation of the E-diagrams (given the elementary vertices and propagators) 
has been shown to hold up to two-loop order for the equal-time two-point 
function and to one-loop order for the four-point function in the context of 
a scalar $\phi^4$ theory \cite{Web09} (a detailed account will be given
elsewhere). Note that two contributions, the second diagram in the 
first line of Fig.\ \ref{figAA} and the third diagram in the last line in the 
same figure corresponding to the first loop integral in Eq.\ \eqref{g2solf2tp}
and to Eq.\ \eqref{g2solf2ct}, respectively, do not fit into this general 
scheme.

The calculation of the equal-time ghost two-point function
from the graphical rules is much simpler. One obtains directly
\be
\langle c^a (\bf{p}_1) \bar{c}^b (\bf{p}_2) \rangle
= \left( \frac{1}{\bf{p}_1^2} + \frac{N_c g^2}{\bf{p}_1^2}
\int \frac{d^3 q}{(2 \pi)^3} \frac{1 - (\hat{\bf{p}}_1 \cdot \hat{\bf{q}})^2}
{(\bf{p}_1 - \bf{q})^2 \, 2 |\bf{q}|} \right) \delta^{ab} \, 
(2 \pi)^3 \delta (\bf{p}_1 + \bf{p}_2) \:, \label{g2gg}
\ee
corresponding to the diagrams in Fig.\ \ref{figcc}.
\begin{figure}
\begin{equation*}
\langle c \bar{c} \rangle =
\parbox{0.96cm}{\begin{center}
\pspicture(-0.11,-0.25)(0.85,0.25)
\psline[linestyle=dashed](0.03,0)(0.71,0)
\psdots[dotstyle=o,dotscale=1.1](0.37,0)
\endpspicture
\end{center}}
+ \parbox{2.4cm}{\begin{center}
\pspicture(-1.2,-0.5)(1.2,0.5)
\SpecialCoor
\multido{\n=17.00+11.25}{15}{%
  \FPadd{-90}{\n}{\m}
  \rput{\m}(0.4;\n){\psCoil{-65}{425}}}
\psline[linestyle=dashed](-0.45,0)(0.45,0)
\psline[linestyle=dashed](-0.45,0)(-1.08,0)
\psline[linestyle=dashed](0.45,0)(1.08,0)
\psdots(-0.45,0)(0.45,0)
\psdots[dotstyle=o,dotscale=1.1](-0.82,0)(0.82,0)(0,0.4)(0,0)
\endpspicture
\end{center}}
\end{equation*}
\caption{Diagrammatic representation of Eq.\ \eqref{g2gg}. \label{figcc}}
\end{figure}
Note that one of the factors $1/\bf{p}_1^2$ for the external ghost
propagators cancels against the momentum dependence of the ghost-gluon 
vertices.

We shall close this section with a calculation of the static (heavy quark)
potential, the energy for a configuration of static external color charges 
$\rho_q (\bf{x})$. To this end, we introduce charges 
$\rho_q (\bf{x})$ into the Hamiltonian, see Eqs.\ \eqref{colch} and 
\eqref{clcoul}. Compared to the
Coulomb term \eqref{cls2}--\eqref{cls3} which was calculated in the absence
of external charges, there are two new terms of order $g^2$:
\be
\frac{g^2}{2} \int \frac{d^3 p}{(2 \pi)^3} \, \rho_q^a (-\bf{p}) 
\frac{1}{\bf{p}^2} \, \rho_q^a (\bf{p}) \:, \label{g2sp}
\ee
which is $\bf{A}$-independent and hence only contributes to the vacuum
energy but leaves the vacuum wave functional unchanged, and
\be
g^2 \int \frac{d^3 p_1}{(2 \pi)^3} \frac{d^3 p_2}{(2 \pi)^3} 
\frac{d^3 p_3}{(2 \pi)^3} \, f^{abc} \, \frac{1}{\bf{p}_1^2} 
(2 \pi)^3 \delta (\bf{p}_1 + \bf{p}_2 + \bf{p}_3) \, \rho^a_q (\bf{p}_1)
A_j^b (\bf{p}_2) \, \frac{1}{i} (2 \pi)^3 
\frac{\delta}{\delta A^c_j (-\bf{p}_3)} \:. \label{clco}
\ee
The latter term, when applied to the vacuum wave functional, generates the
following (properly symmetrized) expression to order $g^2$,
\be
- \frac{g^2}{2} \int \frac{d^3 p_1}{(2 \pi)^3} \frac{d^3 p_2}{(2 \pi)^3} 
\frac{d^3 p_3}{(2 \pi)^3} \, i f^{abc} \, \frac{|\bf{p}_2| - |\bf{p}_3|}
{\bf{p}_1^2} \, \rho^a_q (\bf{p}_1) A_j^b (\bf{p}_2) A_j^c (\bf{p}_3)
(2 \pi)^3 \delta (\bf{p}_1 + \bf{p}_2 + \bf{p}_3) \:, \label{clAAq}
\ee
which implies that the vacuum wave functional to order $g^2$ has to be
modified in order to fulfill the Schr\"odinger equation with this new term.

A term cancelling the expression \eqref{clAAq} in the Schr\"odinger equation
can only result from the second derivative term in the Hamiltonian 
[Eq.\ \eqref{clA2}]. It is then simple to see that we have to add 
the expression
\be
- \frac{g^2}{2} \int \frac{d^3 p_1}{(2 \pi)^3} \frac{d^3 p_2}{(2 \pi)^3} 
\frac{d^3 p_3}{(2 \pi)^3} \, i f^{abc} \, \frac{1}{\bf{p}_1^2} \,
\frac{|\bf{p}_2| - |\bf{p}_3|}{|\bf{p}_2| + |\bf{p}_3|}
\, \rho^a_q (\bf{p}_1) A_j^b (\bf{p}_2) A_j^c (\bf{p}_3)
(2 \pi)^3 \delta (\bf{p}_1 + \bf{p}_2 + \bf{p}_3) \label{g2solAAq}
\ee
to the negative of the exponent of the vacuum wave functional in order to 
satisfy the Schr\"odinger equation to order $g^2$. This term describes the 
back-reaction of the vacuum to the presence of the external charges to order 
$g^2$. Observe that due to the
presence of the external charges $\rho_q (\bf{p})$, the coefficient function 
of $A^a_i (\bf{p}_1) A^b_j (\bf{p}_2)$ in the vacuum wave functional 
ceases to be of the form $f_2 (\bf{p}_2) \delta^{ab} \, \delta_{ij} 
(2 \pi)^3 \delta (\bf{p}_1 + \bf{p}_2)$. Furthermore, contrary
to the terms found before, the contribution \eqref{g2solAAq} is imaginary.
The rest of the vacuum wave functional determined in the previous section 
remains without change.

We now have to calculate the vacuum energy in the presence of the external
charges. To order $g^2$, the result is the former one, Eq.\ \eqref{g2E0},
without any contribution from the new term \eqref{g2solAAq}, plus Eq.\
\eqref{g2sp} which is the part of the energy that depends on the external 
charges and hence defines the potential to this order. Of course, this is 
just the well-known Coulomb potential of electrodynamics. What we are really
interested in are the first quantum corrections to this ``bare'' potential, 
which are of order $g^4$.

In general, the vacuum energy is given by
\be
E_0 = \frac{g^2}{2} \int \frac{d^3 p}{(2 \pi)^3} \,
\rho^a_q (-\bf{p}) \frac{1}{\bf{p}^2} \, \rho^a_q (\bf{p})
- \frac{1}{2} \int \frac{d^3 p}{(2 \pi)^3} \left. (2 \pi)^3 
\frac{\delta}{\delta A_i^a (\bf{p})} (2 \pi)^3
\frac{\delta}{\delta A_i^a (-\bf{p})} \, \psi (\bf{A}) \right|_{\bf{A} = 0}  
\:, \label{xcE0}
\ee
which reduces to Eq.\ \eqref{g2E0} in the absence of external charges.
A contribution of order $g^4$ can hence only originate from the terms in the
vacuum wave functional that are quadratic in $\bf{A}$. As long as we are
only interested in the potential between static sources, we can concentrate
on terms that contain precisely two powers of $\rho_q$. We then start
by identifying all the contributions to the Schr\"odinger equation of order
$g^4$ that contain two powers of $\bf{A}$ and two powers of $\rho_q$.
One of these contributions results from expanding the Coulomb kernel
\be
\langle \bf{x}, a | (-\nabla \cdot \bf{D})^{-1}
(-\nabla^2) (-\nabla \cdot \bf{D})^{-1} | \bf{y}, b \rangle
\ee
in Eq.\ \eqref{christlee} to second order in $\bf{A}$ for $\rho = \rho_q$.
The result is the term
\bmu
- \frac{3}{4} \, g^4 \int \frac{d^3 p_1}{(2 \pi)^3} \cdots 
\frac{d^3 p_4}{(2 \pi)^3} \left( f^{ace} f^{bde} \, \frac{p_{1,i} p_{2,j}}
{\bf{p}_1^2 \, \bf{p}_2^2 \, (\bf{p}_1 + \bf{p}_3)^2} 
+ f^{ade} f^{bce} \, \frac{p_{1,j} p_{2,i}}
{\bf{p}_1^2 \, \bf{p}_2^2 \, (\bf{p}_1 + \bf{p}_4)^2} \right) \\[2mm]
{}\times \rho^a_q (\bf{p}_1) \rho^b_q (\bf{p}_2) A^c_i (\bf{p}_3) 
A^d_j (\bf{p}_4) (2 \pi)^3 \delta (\bf{p}_1 + \ldots + \bf{p}_4) \label{g4xc1}
\emu
on the left-hand side of the Schr\"odinger equation.

Another contribution of the same type arises from the second functional
derivative in Eq.\ \eqref{clA2} acting (twice) on the term \eqref{g2solAAq},
which gives the contribution
\bal
\frac{g^4}{4} \int \frac{d^3 p_1}{(2 \pi)^3} \cdots 
\frac{d^3 p_4}{(2 \pi)^3} \frac{1}{\bf{p}_1^2 \, \bf{p}_2^2} 
&\left( f^{ace} f^{bde} \, t_{ij} (\bf{p}_1 + \bf{p}_3) \,
\frac{|\bf{p}_1 + \bf{p}_3| - |\bf{p}_3|}
{|\bf{p}_1 + \bf{p}_3| + |\bf{p}_3|} \,
\frac{|\bf{p}_1 + \bf{p}_3| - |\bf{p}_4|}
{|\bf{p}_1 + \bf{p}_3| + |\bf{p}_4|} \right. \n \\
&\left. {}+ f^{ade} f^{bce} \, t_{ij} (\bf{p}_1 + \bf{p}_4) \,
\frac{|\bf{p}_1 + \bf{p}_4| - |\bf{p}_3|}
{|\bf{p}_1 + \bf{p}_4| + |\bf{p}_3|} \,
\frac{|\bf{p}_1 + \bf{p}_4| - |\bf{p}_4|}
{|\bf{p}_1 + \bf{p}_4| + |\bf{p}_4|} \right) \n \\[2mm]
&\hspace{1cm} {}\times \rho^a_q (\bf{p}_1) \rho^b_q (\bf{p}_2) A^c_i (\bf{p}_3)
A^d_j (\bf{p}_4) (2 \pi)^3 \delta (\bf{p}_1 + \ldots + \bf{p}_4) \label{g4xc2}
\eal
to the Schr\"odinger equation.
The last contribution of the same type comes from the ``mixed'' term
where the operator \eqref{clco} acts upon the expression 
\eqref{g2solAAq} in the wave functional. The result is
\bal
-\frac{g^4}{4} \int \frac{d^3 p_1}{(2 \pi)^3} \cdots 
\frac{d^3 p_4}{(2 \pi)^3} \frac{1}{\bf{p}_1^2 \, \bf{p}_2^2} 
&\left[ f^{ace} f^{bde} \, t_{ij} (\bf{p}_1 + \bf{p}_3) 
\left( \frac{|\bf{p}_1 + \bf{p}_3| - |\bf{p}_3|}
{|\bf{p}_1 + \bf{p}_3| + |\bf{p}_3|} 
+ \frac{|\bf{p}_1 + \bf{p}_3| - |\bf{p}_4|}
{|\bf{p}_1 + \bf{p}_3| + |\bf{p}_4|} \right) \right. \n \\
&\hspace{-2mm} \left. {}+ f^{ade} f^{bce} \,
t_{ij} (\bf{p}_1 + \bf{p}_4)
\left( \frac{|\bf{p}_1 + \bf{p}_4| - |\bf{p}_3|}
{|\bf{p}_1 + \bf{p}_4| + |\bf{p}_3|} 
+ \frac{|\bf{p}_1 + \bf{p}_4| - |\bf{p}_4|}
{|\bf{p}_1 + \bf{p}_4| + |\bf{p}_4|} \right) \right] \n \\[2mm]
&\hspace{0.7cm} {}\times \rho^a_q (\bf{p}_1) \rho^b_q (\bf{p}_2) A^c_i 
(\bf{p}_3) A^d_j (\bf{p}_4) (2 \pi)^3 \delta (\bf{p}_1 + \ldots + \bf{p}_4) \:,
\label{g4xc3}
\eal
to be included in the Schr\"odinger equation. It can be shown that no 
other contributions quadratic in $\bf{A}$ and in $\rho_q$ exist to order
$g^4$.

In analogy to the determination of the expression \eqref{g2solAAq}
from Eq.\ \eqref{clAAq},
the three contributions \eqref{g4xc1}--\eqref{g4xc3} to the Schr\"odinger 
equation are taken care of by including the following expression in the 
negative exponent of the vacuum wave functional [in addition to \eqref{solf3}, 
\eqref{solf4bv}--\eqref{solf4ct}, \eqref{g2solf2tp}--\eqref{g2solf2ct},
and \eqref{g2solAAq}]
\bmu
-\frac{g^4}{4} \int \frac{d^3 p_1}{(2 \pi)^3} \cdots 
\frac{d^3 p_4}{(2 \pi)^3} \frac{1}{\bf{p}_1^2 \, \bf{p}_2^2 \, 
(|\bf{p}_3| + |\bf{p}_4|)} \left\{ f^{ace} f^{bde} \left[
\frac{3 p_{1,i} p_{2,j}}{(\bf{p}_1 + \bf{p}_3)^2} 
- t_{ij} (\bf{p}_1 + \bf{p}_3) \right. \right. \\
\left. {}\times \left(
\frac{|\bf{p}_1 + \bf{p}_3| - |\bf{p}_3|}
{|\bf{p}_1 + \bf{p}_3| + |\bf{p}_3|} \, 
\frac{|\bf{p}_1 + \bf{p}_3| - |\bf{p}_4|}
{|\bf{p}_1 + \bf{p}_3| + |\bf{p}_4|} 
- \frac{|\bf{p}_1 + \bf{p}_3| - |\bf{p}_3|}
{|\bf{p}_1 + \bf{p}_3| + |\bf{p}_3|} 
- \frac{|\bf{p}_1 + \bf{p}_3| - |\bf{p}_4|}
{|\bf{p}_1 + \bf{p}_3| + |\bf{p}_4|} \right) \right] \\
{}+ f^{ade} f^{bce} \left[ \frac{3 p_{1,j} p_{2,i}}
{(\bf{p}_1 + \bf{p}_4)^2} - t_{ij} (\bf{p}_1 + \bf{p}_4)
\left( \frac{|\bf{p}_1 + \bf{p}_4| - |\bf{p}_3|}
{|\bf{p}_1 + \bf{p}_4| + |\bf{p}_3|} \,
\frac{|\bf{p}_1 + \bf{p}_4| - |\bf{p}_4|}
{|\bf{p}_1 + \bf{p}_4| + |\bf{p}_4|} 
- \frac{|\bf{p}_1 + \bf{p}_4| - |\bf{p}_3|}
{|\bf{p}_1 + \bf{p}_4| + |\bf{p}_3|} \right. \right. \\
\left. \left. \left. {}- \frac{|\bf{p}_1 + \bf{p}_4| - |\bf{p}_4|}
{|\bf{p}_1 + \bf{p}_4| + |\bf{p}_4|} \right) \right] \right\}
\rho^a_q (\bf{p}_1) \rho^b_q (\bf{p}_2) A^c_i (\bf{p}_3) 
A^d_j (\bf{p}_4) (2 \pi)^3 \delta (\bf{p}_1 + \ldots + \bf{p}_4) \:.
\label{g4solAAqq}
\emu
This result (multiplied by 2) is represented 
diagrammatically in Fig.\ \ref{figAArr}, where we have denoted the
``Coulomb propagator'' $1/\bf{p}^2$ as a double line.
\begin{figure}
\begin{equation*}
- 3 \bigg( \parbox{1.6cm}{\begin{center}
\pspicture(-0.4,-0.4)(0.97,0.4)
\pscoil[coilarmA=0.1,coilarmB=0](0.57,0)(0.95,0.38)
\psline[doubleline=true,linewidth=0.8pt,doublesep=0.6pt](0.57,0)(0.95,-0.38)
\psline[doubleline=true,linewidth=0.8pt,doublesep=0.6pt](0,0)(0.57,0)
\pscoil[coilarmA=0.1,coilarmB=0](0,0)(-0.38,0.38)
\psline[doubleline=true,linewidth=0.8pt,doublesep=0.6pt](0,0)(-0.38,-0.38)
\psdots(0,0)(0.57,0)
\endpspicture
\end{center}}
+ \text{1 perm.} \bigg)
- 2 \bigg( \parbox{1.6cm}{\begin{center}
\pspicture(-0.4,-0.4)(0.97,0.4)
\pscoil[coilarmA=0.1,coilarmB=0](0.57,0)(0.95,0.38)
\psline[doubleline=true,linewidth=0.8pt,doublesep=0.6pt](0.57,0)(0.95,-0.38)
\pscoil[coilarm=0.1](0,0)(0.57,0)
\pscoil[coilarmA=0.1,coilarmB=0](0,0)(-0.38,0.38)
\psline[doubleline=true,linewidth=0.8pt,doublesep=0.6pt](0,0)(-0.38,-0.38)
\psdots(0,0)(0.57,0)
\endpspicture
\end{center}}
+ \text{1 perm.} \bigg)
\end{equation*}
\caption{Diagrammatic representation of the contributions \eqref{g4solAAqq}
(multiplied by 2) to the vacuum wave functional. \label{figAArr}}
\end{figure}
The first diagram (and its
permutation) corresponds to the expression \eqref{g4xc1}, while the second
diagram (plus its permutation) corresponds to the sum of the expressions
\eqref{g4xc2} and \eqref{g4xc3}. From Eq.\ \eqref{xcE0}, we find the 
contribution to the vacuum energy
\bmu
\frac{g^2}{2} \int \frac{d^3 p}{(2 \pi)^3} \, \rho^a_q (-\bf{p})
\frac{1}{(\bf{p}^2)^2} \left\{ \frac{N_c g^2}{2} 
\int \frac{d^3 q}{(2 \pi)^3} \frac{1}{2 |\bf{q}|} \left[ 
3 \, \frac{\bf{p}^2 - (\bf{p} \cdot \hat{\bf{q}})^2}{(\bf{p} - \bf{q})^2} 
\right. \right. \\
{}+ t_{ij} (\bf{q}) t_{ij} (\bf{p} - \bf{q}) \left( 
\left( \frac{|\bf{p} - \bf{q}| - |\bf{q}|}{|\bf{p} - \bf{q}| + |\bf{q}|}
\right)^2 - 2 \, \frac{|\bf{p} - \bf{q}| - |\bf{q}|}
{|\bf{p} - \bf{q}| + |\bf{q}|} \right)
+ 3 \, \frac{\bf{p}^2 - (\bf{p} \cdot \hat{\bf{q}})^2}{(\bf{p} + \bf{q})^2} \\ 
\left. \left. {}+ t_{ij} (\bf{q}) t_{ij} (\bf{p} + \bf{q}) \left(
\left( \frac{|\bf{p} + \bf{q}| - |\bf{q}|}{|\bf{p} + \bf{q}| + |\bf{q}|}
\right)^2 - 2 \, \frac{|\bf{p} + \bf{q}| - |\bf{q}|}
{|\bf{p} + \bf{q}| + |\bf{q}|} \right) \right] \right\} 
\rho_q^a (\bf{p}) \:.
\emu
This latter expression can be simplified by shifting 
$\bf{q} \to \bf{q} - \bf{p}$ in the last two terms (in the round 
bracket). Together with Eq.\ \eqref{g2sp}, we find for the part of the vacuum 
energy that is quadratic in the external static charge $\rho_q$,
\be
E_0^{(\rho_q, 2)} =
\frac{g^2}{2} \int \frac{d^3 p}{(2 \pi)^3} \, \rho^a_q (-\bf{p}) V (\bf{p})
\rho_q^a (\bf{p}) \:, \label{defsp}
\ee
the following result for the static potential to order $g^2$
\bal
V (\bf{p}) &= \frac{1}{\bf{p}^2} + \frac{N_c g^2}{\bf{p}^2} \, 
3 \int \frac{d^3 q}{(2 \pi)^3} \frac{1 - (\hat{\bf{p}} \cdot \hat{\bf{q}})^2}
{2 |\bf{q}| \, (\bf{p} - \bf{q})^2} \label{g4spat} \\
&\hspace{1.7cm} {}- \frac{N_c g^2}{(\bf{p}^2)^2} \int \frac{d^3 q}{(2 \pi)^3}
\left( 1 + \frac{\big( (\bf{p} - \bf{q}) \cdot 
\bf{q} \big)^2}{(\bf{p} - \bf{q})^2 \, \bf{q}^2} \right)
\frac{(|\bf{p} - \bf{q}| - |\bf{q}|)^2}
{2 |\bf{p} - \bf{q}| \, 2 |\bf{q}| \, (|\bf{p} - \bf{q}| + |\bf{q}|)}
\:. \label{g4spst}
\eal
Note that to the proper color Coulomb potential (see, e.g., Ref.\ 
\cite{Zwa98}), only the antiscreening term, the integral in 
Eq.\ \eqref{g4spat}, contributes, while the full static potential also 
contains the screening contribution \eqref{g4spst}. The semi-analytical 
variational approaches \cite{FR04a,FR04b,RF05,ERS07,ERS08} have only
considered the proper color Coulomb potential so far.

We can associate diagrams with the different contributions in
Eqs.\ \eqref{g4spat}--\eqref{g4spst} in a natural way. The vertex that joins 
two Coulomb (double) lines and one gluon line corresponds to the same
mathematical expression as the ghost-gluon vertex since both objects 
originate from the Faddeev-Popov operator $(-\nabla \cdot \bf{D})$ (or its 
inverse). On the other hand, the vertex with two gluon lines and one Coulomb 
line translates to the expression
\be
i g f^{abc} \, \frac{|\bf{p}_2| - |\bf{p}_3|}{|\bf{p}_2| + |\bf{p}_3|}
\, \delta_{jk} \:,
\ee
where the gluon lines carry the momenta $\bf{p}_2$ and $\bf{p}_3$, (spatial)
Lorentz indices $j$ and $k$, and color indices $b$ and $c$ [cf.\ Eq.\ 
\eqref{g2solAAq}]. Note that the ``elementary'' Coulomb interaction 
\eqref{solf4ct} is \emph{different} from the contraction of two such vertices 
with a Coulomb propagator. With these conventions, we can represent the static
potential as in Fig.\ \ref{figV}. The E-operator in Fig.\ \ref{figV}
exclusively refers to the internal gluon propagators and thus amounts to
multiplying the integrand with $|\bf{p} - \bf{q}| + |\bf{q}|$.
\begin{figure}
\begin{equation*}
V = 
%
\parbox{0.96cm}{\begin{center}
\pspicture(-0.11,-0.25)(0.85,0.25)
\psline[doubleline=true,linewidth=0.8pt,doublesep=0.6pt](0.03,0)(0.71,0)
\endpspicture
\end{center}}
+ 3 \parbox{2.2cm}{\begin{center}
\pspicture(-1.05,-0.5)(1.05,0.5)
\SpecialCoor
\multido{\n=17.00+11.25}{15}{%
  \FPadd{-90}{\n}{\m}
  \rput{\m}(0.4;\n){\psCoil{-65}{425}}}
\psline[doubleline=true,linewidth=0.8pt,doublesep=0.6pt](-0.45,0)(0.45,0)
\psline[doubleline=true,linewidth=0.8pt,doublesep=0.6pt](-0.45,0)(-0.92,0)
\psline[doubleline=true,linewidth=0.8pt,doublesep=0.6pt](0.45,0)(0.92,0)
\psdots(-0.45,0)(0.45,0)
\psdots[dotstyle=o,dotscale=1.1](0,0.4)
\endpspicture
\end{center}}
+ 2 E \bigg( \parbox{1.9cm}{\begin{center}
\pspicture(-0.95,-0.5)(0.95,0.5)
\SpecialCoor
\multido{\n=17.00+11.25}{15}{%
  \FPadd{-90}{\n}{\m}
  \rput{\m}(0.4;\n){\psCoil{-65}{425}}}
\multido{\n=197.00+11.25}{15}{%
  \FPadd{-90}{\n}{\m}
  \rput{\m}(0.4;\n){\psCoil{-65}{425}}}
\psline[doubleline=true,linewidth=0.8pt,doublesep=0.6pt](-0.45,0)(-0.92,0)
\psline[doubleline=true,linewidth=0.8pt,doublesep=0.6pt](0.45,0)(0.92,0)
\psdots(-0.45,0)(0.45,0)
\psdots[dotstyle=o,dotscale=1.1](0,0.4)(0,-0.4)
\endpspicture
\end{center}} \bigg)
\end{equation*}
\caption{A diagrammatic interpretation of the static potential to order
$g^2$. \label{figV}}
\end{figure}

We hence have succeeded in calculating the equal-time gluon and ghost
two-point functions and the static potential to one-loop order in our
functional perturbative approach, with the results \eqref{g2AAtp}--\eqref{g2gg}
and \eqref{g4spat}--\eqref{g4spst}. The same results can be obtained from a 
straightforward application of Rayleigh-Schr\"odinger perturbation theory 
\cite{CRW09}. Compared to these latter calculations, we have here
developed a functional integral and diagrammatical approach that is
potentially advantageous in higher-order perturbative calculations. We
have also described a set of simplified diagrammatic rules (the 
``E-operator'') for the determination of equal-time correlation functions
that is expected to carry over to higher perturbative orders and is
hoped to eventually lead to nonperturbative equations for the equal-time 
correlation functions analogous to Dyson-Schwinger equations.

\section{Lagrangian approach and renormalization}

Naive power counting shows that the results of the preceding section,
Eqs.\ \eqref{g2AAtp}--\eqref{g2gg} and \eqref{g4spat}--\eqref{g4spst}, are 
ultraviolet (UV) divergent and need to be renormalized. However, some of the 
denominators occuring in the loop integrals are of a different
type from those that usually appear in covariant perturbation theory, and
efficient techniques for the handling of these terms have yet to be
developed. These remarks apply in particular to Eqs.\ \eqref{g2AAgl} and
\eqref{g4spst}.

The equal-time correlation functions we have been calculating are a
special or limiting case of the usual space-time correlation functions, whence
we naturally obtain the representation
\be
\langle A_i^a (\bf{p}_1, t=0) A_j^b (\bf{p}_2, t=0) \rangle =
\int_{-\infty}^\infty \frac{d p_{1,4}}{2 \pi} \frac{d p_{2,4}}{2 \pi}
\langle A_i^a (\bf{p}_1, p_{1,4}) A_j^b (\bf{p}_2, p_{2,4}) \rangle 
\label{etreduct}
\ee
(with the space-time correlation functions written in Euclidean space-time). 
Since the regularization and renormalization program has been
developed for the space-time correlation functions, this representation
is quite useful for our purposes. In the case of Coulomb gauge Yang-Mills
theory, however, covariance is explicitly broken through the gauge
condition, and the calculation of the space-time correlation functions
in the usual Lagrangian functional integral approach
represents a difficulty by itself. Techniques have been developed to
overcome these difficulties and applied in Ref.\ \cite{WR07b}
to the calculation of the two-point correlation functions to one-loop
order.

Before properly considering the renormalization of our results, we will
verify that these coincide on a formal level with the expressions 
obtained via Eq.\ \eqref{etreduct}, taking for the space-time correlation 
functions the formulas derived in the Lagrangian functional integral approach
in Ref.\ \cite{WR07b}. In our notation, 
\bal
\lefteqn{\langle A_i^a (p_1) A_j^b (p_2) \rangle 
= \left( \frac{1}{p_1^2} - \frac{N_c g^2}{\big( p_1^2 \big)^2} \, 
\frac{4}{3} \int \frac{d^4 q}{(2 \pi)^4} \frac{1}{q^2} \right.} \hspace{1.7cm} 
\label{covAAtp} \\
&\phantom{=} {}+ \frac{N_c g^2}{\big( p_1^2 \big)^2} 
\int \frac{d^4 q}{(2 \pi)^4} 
\frac{\big( \delta_{km} p_{1,n} + \delta_{mn} q_k - \delta_{nk} p_{1,m} \big) 
\, t_{mr} (\bf{p}_1 - \bf{q}) \, t_{ns} (\bf{q})}
{q^2} \n \\
&\phantom{=} \hspace{3cm} {}\times 
\frac{\big( \delta_{lr} p_{1,s} + \delta_{rs} q_l 
- \delta_{sl} p_{1,r} \big) \, t_{kl} (\bf{p}_1)}{(p_1 - q)^2} \\
&\phantom{=} \left. {}+ \frac{N_c g^2}{\big( p_1^2 \big)^2} \,
\frac{1}{2} \int \frac{d^4 q}{(2 \pi)^4} \frac{t_{kl} (\bf{p}_1)
t_{kl} (\bf{q}) \, (p_{1,4}^2 - \bf{q}^2)}
{q^2 \, (\bf{p}_1 - \bf{q})^2} \right) \delta^{ab} \, 
t_{ij} (\bf{p}_1) (2 \pi)^4 \delta (p_1 + p_2) \:, \label{covAAct}
\eal
with $p_1^2 \equiv \bf{p}_1^2 + p_{1,4}^2$ in Euclidean space-time.
Formal integration of $p_{1,4}$ and $p_{2,4}$ as in Eq.\ 
\eqref{etreduct} and of the component $q_4$ of the loop momentum, most
easily using the residue theorem, leads to our equal-time correlation function 
\eqref{g2AAtp}--\eqref{g2AAct}. In Eq.\ \eqref{covAAtp}, we have included the
tadpole diagram in order that the correspondence with the equal-time gluon
two-point function \eqref{g2AAtp}--\eqref{g2AAct} be term by term.
The tadpole diagram was not considered explicitly in Refs.\ \cite{WR07b,WR08}
because it vanishes in dimensional regularization.

The case of the ghost two-point function is even simpler, because it is
instantaneous already in the Lagrangian approach: from Ref.\ \cite{WR07b}, 
\be
\langle c^a (p_1) \bar{c}^b (p_2) \rangle
= \left( \frac{1}{\bf{p}_1^2} + \frac{N_c g^2}{\big( \bf{p}_1^2 \big)^2}
\int \frac{d^4 q}{(2 \pi)^4} \frac{p_{1,i} p_{1,j} \, 
t_{ij} (\bf{q})}{q^2 \, (\bf{p}_1 - \bf{q})^2} \right) 
\delta^{ab} \, (2 \pi)^4 \delta (p_1 + p_2) \:, \label{covgg}
\ee
and the fact that the dependence on $p_{1,4}$ and $p_{2,4}$ is exclusively 
through the delta function for energy conservation implies that
$\langle c^a (\bf{p}_1, t_1) \bar{c}^b (\bf{p}_2, t_2) \rangle$ contains
the factor $\delta(t_1 - t_2)$. Integrating over $p_{1,4}$ \emph{and} $p_{2,4}$
or putting $t_1$ \emph{and} $t_2$ to zero then results in a factor
$\delta (0)$. In this case, in order to reproduce Eq.\ \eqref{g2gg}, we
integrate \emph{either} over $p_{1,4}$ \emph{or} over $p_{2,4}$, which just
eliminates the delta function for energy conservation [more symmetrically,
one may integrate over $(p_{1,4} + p_{2,4})$ instead]. Performing the 
integral over $q_4$ then converts Eq.\ \eqref{covgg} to Eq.\ \eqref{g2gg}.

Although it is certainly not surprising that the Lagrangian functional integral
approach of Refs.\ \cite{WR07a,WR07b,WR08} gives the same results for
the equal-time correlation functions as our Hamiltonian approach, it is also
not trivial. Concerning the gauge fixing procedure, there is the 
following important difference between the two approaches: in
the Lagrangian formulation, the Weyl gauge $A_0 \equiv 0$ cannot be 
implemented in addition to the Coulomb gauge condition 
$\nabla \cdot \bf{A} \equiv 0$ \cite{WR07c}. Indeed, in the 
first-order Lagrangian formalism, integrating out the $A_0$-field rather 
than setting $A_0$ to zero yields an expression in the exponent of the measure 
for the functional integral that resembles the Christ-Lee 
Hamiltonian \cite{WR07a}. The derivation of the Hamilton operator 
\eqref{christlee} by Christ and Lee \cite{CL80}, on the other hand, relies on 
the existence of a gauge transformation that makes any gauge field $A_\mu$ 
satisfy both the Coulomb and Weyl gauge conditions. We discuss
the possibility of simultaneously implementing the Weyl and Coulomb gauges 
in the Hamiltonian and the Lagrangian approaches in the Appendix.
 
Although not defined as an equal-time correlation function in our approach, 
it turns out that the static potential \eqref{g4spat}--\eqref{g4spst} is 
related to the space-time two-point function $\langle A_0^a (p_1) A_0^b (p_2) 
\rangle$ in the Lagrangian functional integral approach, 
as was first pointed out by Zwanziger \cite{Zwa98}. The formal
expression for the space-time correlation function is \cite{WR07b}
\bal
\lefteqn{ \langle A_0^a (p_1) A_0^b (p_2) \rangle
= \left( \frac{1}{\bf{p}_1^2} + \frac{N_c g^2}{\big( \bf{p}_1^2 \big)^2} \, 
3 \int \frac{d^4 q}{(2 \pi)^4} \frac{p_{1,i} p_{1,j} \, t_{ij} (\bf{q})}
{q^2 \, (\bf{p}_1 - \bf{q})^2} \right.} \hspace{1.6cm} \label{covspat} \\
&{}+ \left. \frac{N_c g^2}{\big( \bf{p}_1^2 \big)^2} 
\int \frac{d^4 q}{(2 \pi)^4} 
\frac{q_4}{p_{1,4}} \, \frac{\bf{p}_1 \cdot (\bf{p}_1 - 2 \bf{q})}
{q^2 \, (p_1 - q)^2} \, t_{ij} (\bf{p}_1 - \bf{q})
t_{ij} (\bf{q}) \right) \delta^{ab} \, (2 \pi)^4 \delta (p_1 + p_2)
\:. \label{covspst}
\eal
Integrating over either $p_{1,4}$ or $p_{2,4}$ and over the energy
component $q_4$ of the loop momentum, we obtain the antiscreening
contribution \eqref{g4spat} to the static potential from Eq.\ \eqref{covspat}
because the latter is already instantaneous. 
In order to find Eq.\ \eqref{g4spst} starting from Eq.\
\eqref{covspst}, we have to put the respective other energy component,
$p_{2,4}$ or $p_{1,4}$, to zero in addition [this is not necessary in the
cases of Eqs.\ \eqref{covgg} and \eqref{covspat}, because there the result of 
integrating over one of the energy components is independent of the other]. 
For $\langle A_0^a (\bf{p}_1, t_1) A_0^b (\bf{p}_2, t_2) \rangle$, this 
procedure amounts to integrating over the relative time $t_1 - t_2$, which is,
in fact, intuitively quite appealing for a non-instantaneous contribution
to the potential between static sources.

We shall now use the representation \eqref{etreduct} of the equal-time
gluon two-point function and the corresponding representations of
the ghost two-point function and the static potential for the renormalization
of these equal-time correlation functions. To this end, we 
make use of the explicit expressions obtained for Eqs.\ 
\eqref{covAAtp}--\eqref{covspst} in
Ref.\ \cite{WR07b} in dimensional regularization. Thus, by integrating
the result for \eqref{covAAtp}--\eqref{covAAct} according to Eq.\ 
\eqref{etreduct}, we obtain for the equal-time correlation function
\be
\langle A_i^a (\bf{p}_1) A_j^b (\bf{p}_2) \rangle = 
\left[ \frac{1}{2 |\bf{p}_1|} + \frac{N_c g^2}{(4 \pi)^2} \, 
\frac{1}{2 |\bf{p}_1|} \left( \frac{1}{\epsilon} - \ln \frac{\bf{p}_1^2}{\mu^2}
+ C_A \right) \right] \delta^{ab} \, t_{ij} (\bf{p}_1) 
(2 \pi)^3 \delta (\bf{p}_1 + \bf{p}_2) \label{regAA}
\ee
in the limit $\epsilon \to 0$, where $d = 3 - 2 \epsilon$ is the dimension
of space and $\mu$ an arbitrary mass scale. The value of the constant
$C_A$ is not relevant to our purposes, but being an integral over an
explicitly known function of $p_{1,4}^2/\bf{p}_1^2$, we have carefully
checked that it is finite.

From the explicit expressions for Eqs.\ \eqref{covgg}--\eqref{covspst} in
dimensional regularization \cite{WR07b}, we find directly
\bal
\langle c^a (\bf{p}_1) \bar{c}^b (\bf{p}_2) \rangle &= 
\left[ \frac{1}{\bf{p}_1^2} + \frac{N_c g^2}{(4 \pi)^2} \, 
\frac{1}{\bf{p}_1^2} \, \frac{4}{3} \left( \frac{1}{\epsilon} - 
\ln \frac{\bf{p}_1^2}{\mu^2} + C_c \right) \right] \delta^{ab} \, 
(2 \pi)^3 \delta (\bf{p}_1 + \bf{p}_2) \:, \label{reggg} \\
V (\bf{p}_1) &= \frac{1}{\bf{p}_1^2} + \frac{N_c g^2}{(4 \pi)^2} \, 
\frac{1}{\bf{p}_1^2} \, \frac{11}{3} \left( \frac{1}{\epsilon} - 
\ln \frac{\bf{p}_1^2}{\mu^2} + C_V \right) \:. \label{regsp}
\eal

This procedure to regularize the equal-time correlation functions finds
further support in the cases where the equal-time functions in the form 
\eqref{g2AAtp}--\eqref{g2gg} and \eqref{g4spat}--\eqref{g4spst} can be
evaluated directly in dimensional regularization (in $d = 3 - 2 \epsilon$
dimensions). For Eqs.\ \eqref{g2AAct} and \eqref{g2gg}, identical results are
obtained in both ways \cite{CRW09} [also trivially for Eq.\ 
\eqref{g2AAtp} and the loop integral in Eq.\ \eqref{g4spat} which is just
three times the one of Eq.\ \eqref{g2gg}].

The results \eqref{regAA} and \eqref{reggg} for the equal-time two-point
correlation functions can be renormalized in analogy to the procedures
developed for covariant theories: we introduce renormalized correlation
functions (or correlation functions of the renormalized fields)
\bal
\langle A_{R,i}^a (\bf{p}_1) A_{R,j}^b (\bf{p}_2) \rangle &=
\frac{1}{Z_A} \langle A_i^a (\bf{p}_1) A_j^b (\bf{p}_2) \rangle \:, \n \\
\langle c_R^a (\bf{p}_1) \bar{c}_R^b (\bf{p}_2) \rangle &=
\frac{1}{Z_c} \langle c^a (\bf{p}_1) \bar{c}^b (\bf{p}_2) \rangle \:.
\label{defmultren}
\eal
The simplest choice of the normalization conditions is
\bal
\left. \langle A_{R,i}^a (\bf{p}_1) A_{R,j}^b (\bf{p}_2) \rangle 
\right|_{\bf{p}_1^2 = \kappa^2} &= \frac{1}{2 |\bf{p}_1|} \,
\delta^{ab} \, t_{ij} (\bf{p}_1) 
(2 \pi)^3 \delta (\bf{p}_1 + \bf{p}_2) \:, \n \\
\left. \langle c_R^a (\bf{p}_1) \bar{c}_R^b (\bf{p}_2) \rangle
\right|_{\bf{p}_1^2 = \kappa^2} &= \frac{1}{\bf{p}_1^2} \,
\delta^{ab} \, (2 \pi)^3 \delta (\bf{p}_1 + \bf{p}_2) \:, \label{rencond}
\eal
at the renormalization scale $\kappa$. With these normalization conditions
and the results \eqref{regAA}, \eqref{reggg}, we obtain
\bal
Z_A (\kappa) &= 1 + \frac{N_c g^2}{(4 \pi)^2} \left( \frac{1}{\epsilon} - 
\ln \frac{\kappa^2}{\mu^2} + C_A \right) \:, \n \\
Z_c (\kappa) &= 1 + \frac{N_c g^2}{(4 \pi)^2} \, \frac{4}{3} 
\left( \frac{1}{\epsilon} - \ln \frac{\kappa^2}{\mu^2} + C_c \right) \:,
\label{wfr}
\eal
to order $g^2$.

The expression \eqref{regsp} for the static potential needs to be
renormalized, too. This is most naturally achieved by a 
renormalization of the coupling constant as was first suggested in Ref.\ 
\cite{Dre81,Lee81}, through
\be
\left. g^2 V (\bf{p}) \right|_{\bf{p}^2 = \kappa^2} 
= \frac{\bar{g}_R^2 (\kappa)}{\bf{p}^2} \:, \label{defgbar}
\ee
see Eq.\ \eqref{defsp}. Hence, from Eq.\ \eqref{regsp},
\be\label{gbar}
\bar{g}_R^2 (\kappa) = g^2 \left[ 1 + \frac{N_c g^2}{(4 \pi)^2} \, 
\frac{11}{3} \left( \frac{1}{\epsilon} - \ln \frac{\kappa^2}{\mu^2} + C_V 
\right) \right] \:,
\ee
which implies for the corresponding beta function to one-loop order,
\be
\kappa^2 \frac{\d}{\d \kappa^2} \, \bar{g}_R^2 (\kappa) =
\frac{\bar{\beta}_0}{(4 \pi)^2} \, \bar{g}_R^4 (\kappa) \:, \label{defbeta}
\ee
that
\be
\bar{\beta}_0 = -\frac{11}{3} \, N_c \:. \label{g2beta}
\ee
This is the well-known result from covariant perturbation theory (for
Yang-Mills theory in covariant gauges), and has also been found in Ref.\ 
\cite{WR07b}.

For the rest of this section, we will pursue a more conventional way of
renormalizing the coupling constant (which, however, leads to the same
result). To this end, we consider the equal-time ghost-gluon three-point
correlation function
$\langle c^a (\bf{p}_1) \bar{c}^b (\bf{p}_2) A^c_i (\bf{p}_3) \rangle$
to order $g^3$ (one loop). The calculation 
of this correlation function is performed in analogy with 
the determination of the equal-time two-point correlation functions in
Section 3, using the result for the vacuum wave functional obtained in
Section 2. In this particularly simple case (and to the order considered), the
external ``propagators'' (equal-time two-point functions) can be factorized
to define the equal-time proper three-point vertex 
$\Gamma^{abc}_i (\bf{p}_1, \bf{p}_2, \bf{p}_3)$ as
\bmu
\langle c^a (\bf{p}_1) \bar{c}^b (\bf{p}_2) A^c_i (\bf{p}_3) \rangle
= - \int \frac{d^3 p_4}{(2 \pi)^3} \frac{d^3 p_5}{(2 \pi)^3} 
\frac{d^3 p_6}{(2 \pi)^3} \, 
\langle c^a (\bf{p}_1) \bar{c}^d (-\bf{p}_4) \rangle \\
{}\times \Gamma^{def}_j (\bf{p}_4, \bf{p}_5, \bf{p}_6) \,
\langle c^e (-\bf{p}_5) \bar{c}^b (\bf{p}_2) \rangle \,
\langle A_j^f (-\bf{p}_6) A_i^c (\bf{p}_3) \rangle \:.
\emu

The explicit perturbative result is
\bal
\lefteqn{\Gamma^{abc}_j (\bf{p}_1, \bf{p}_2, \bf{p}_3) = -i g f^{abc} 
\Bigg( p_{1,k} - \frac{N_c g^2}{2} \int \frac{d^3 q}{(2 \pi)^3} 
\frac{\big[ \bf{p}_1 \cdot \bf{p}_2 - (\bf{p}_1 \cdot \hat{\bf{q}}) 
(\bf{p}_2 \cdot \hat{\bf{q}}) \big] (p_{1,k} - q_k)}{2 |\bf{q}| \, 
(\bf{p}_1 - \bf{q})^2 \, (\bf{p}_2 + \bf{q})^2}} \hspace{2cm} 
\label{g3ggv1} \\
&{}+ \frac{2 N_c g^2}{2} \int \frac{d^3 q}{(2 \pi)^3}
\frac{p_{1,l} p_{2,n} \, t_{lm} (\bf{p}_1 - \bf{q})
t_{nr} (\bf{p}_2 + \bf{q})}
{\bf{q}^2 \, 2 |\bf{p}_1 - \bf{q}| \, 2 |\bf{p}_2 + \bf{q}|} \n \\
&\phantom{+} {}\times \frac{\delta_{km} (p_{1,r} - p_{3,r} - q_r)
- \delta_{mr} (p_{1,k} - p_{2,k} - 2 q_k) - \delta_{rk} (p_{2,m} - p_{3,m}
+ q_m)}{|\bf{q}| + |\bf{p}_1 - \bf{q}| + |\bf{p}_2 + \bf{q}|} \Bigg) 
\label{g3ggv2} \\[2mm]
&\hspace{6cm} {}\times t_{jk} (\bf{p}_3) (2 \pi)^3
\delta (\bf{p}_1 + \bf{p}_2 + \bf{p}_3) \:. \n 
\eal
It is represented diagrammatically in Fig.\ \ref{figGamma}.
\begin{figure}
\begin{equation*}
\Gamma = - \parbox{1.0cm}{\begin{center}
\pspicture(-0.5,-0.5)(0.5,0.6)
\psline[linestyle=dashed](0,0)(-0.38,-0.38)
\psline[linestyle=dashed](0,0)(0.38,-0.38)
\pscoil[coilarmA=0.1,coilarmB=0](0,0)(0,0.47)
\psdots(0,0)
\endpspicture
\end{center}}
- \parbox{1.6cm}{\begin{center}
\pspicture(-0.8,-0.6)(0.8,0.7)
\psline[linestyle=dashed](-0.37,-0.18)(0,0.19)
\psline[linestyle=dashed](0.37,-0.18)(0,0.19)
\pscoil[coilarm=0.1](-0.37,-0.18)(0.37,-0.18)
\psline[linestyle=dashed](-0.37,-0.18)(-0.68,-0.49)
\psline[linestyle=dashed](0.37,-0.18)(0.68,-0.49)
\pscoil[coilarmA=0.1,coilarmB=0](0,0.19)(0,0.58)
\psdots(-0.37,-0.18)(0.37,-0.18)(0,0.19)
\psdots[dotstyle=o,dotscale=1.1](0,-0.18)(-0.185,0.005)(0.185,0.005)
\endpspicture
\end{center}}
- \parbox{1.7cm}{\begin{center}
\pspicture(-0.85,-0.64)(0.85,0.73)
\pscoil[coilarm=0.1](-0.405,-0.2)(0,0.205)
\pscoil[coilarm=0.1](0.405,-0.2)(0,0.205)
\psline[linestyle=dashed](-0.405,-0.2)(0.405,-0.2)
\psline[linestyle=dashed](-0.405,-0.2)(-0.715,-0.51)
\psline[linestyle=dashed](0.405,-0.2)(0.715,-0.51)
\pscoil[coilarmA=0.1,coilarmB=0](0,0.205)(0,0.595)
\psdots(-0.405,-0.2)(0.405,-0.2)(0,0.205)
\psdots[dotstyle=o,dotscale=1.1](0,-0.2)(-0.203,0.003)(0.203,0.003)
\endpspicture
\end{center}}
\end{equation*}
\caption{The proper ghost-gluon vertex to one-loop order. \label{figGamma}}
\end{figure}
Note that due to the transversality of the gauge, two powers of the external 
momenta can be factorized from the loop integrals [cf.\ Eq.\ 
\eqref{g2gg}] and, as a result, the integrals are UV finite. This phenomenon 
is well-known in another transverse gauge, the Landau gauge \cite{Tay71,MP78}.
For future use, we note that by very lengthy algebra the tensor 
structure in Eq.\ \eqref{g3ggv2} can be simplified as follows:
\bal
&\phantom{=} p_{1,l} \, p_{2,n} \, t_{lm} (\bf{p}_1 - \bf{q})
t_{nr} (\bf{p}_2 + \bf{q}) \Big[ \delta_{km} (p_{1,r} - p_{3,r} - q_r) 
\n \\[1mm]
&\phantom{=} \hspace{1cm} {}- \delta_{mr} (p_{1,k} - p_{2,k} - 2 q_k) - 
\delta_{rk} (p_{2,m} - p_{3,m} + q_m) \Big] t_{jk} (\bf{p}_3) (2 \pi)^3
\delta (\bf{p}_1 + \bf{p}_2 + \bf{p}_3) \n \\
&= 2 \, \Bigg( \bf{q}^2 p_{1,k} + (\bf{p}_1 \cdot \bf{p}_2) q_k 
- \frac{(\bf{p}_1 - \bf{q}) \cdot (\bf{p}_2 + \bf{q})}
{(\bf{p}_1 - \bf{q})^2 \, (\bf{p}_2 + \bf{q})^2} \Big\{ 
[\bf{q} \cdot (\bf{p}_1 - \bf{q})] [\bf{q} \cdot (\bf{p}_2 + \bf{q})] p_{1,k}
\n \\
&\phantom{=} \hspace{3.2cm} {}+ [\bf{p}_1 \cdot (\bf{p}_1 - \bf{q})] 
[\bf{p}_2 \cdot (\bf{p}_2 + \bf{q})] q_k \Big\} \Bigg) \,
t_{jk} (\bf{p}_3) (2 \pi)^3 \delta (\bf{p}_1 + \bf{p}_2 + \bf{p}_3) \:.
\eal

We define the renormalized coupling constant in analogy with the covariant 
case as
\bal
\Gamma^{abc}_{R,j} (\bf{p}_1, \bf{p}_2, \bf{p}_3) 
\Big|_{\bf{p}_1^2 = \bf{p}_2^2 = \bf{p}_3^2 = \kappa^2} 
&\equiv Z_c (\kappa) Z_A^{1/2} (\kappa) \, 
\Gamma^{abc}_j (\bf{p}_1, \bf{p}_2, \bf{p}_3) 
\Big|_{\bf{p}_1^2 = \bf{p}_2^2 = \bf{p}_3^2 = \kappa^2} \n \\
&= -i g_R (\kappa) f^{abc} \, p_{1,k} \, t_{jk} (\bf{p}_3) (2 \pi)^3
\delta (\bf{p}_1 + \bf{p}_2 + \bf{p}_3) \label{grendef}
\eal
at the symmetric point. As a consequence, using Eq.\ \eqref{wfr}
and the UV finiteness of the loop integrals \eqref{g3ggv1}--\eqref{g3ggv2},
\be
g_R (\kappa) = g \left[ 1 + \frac{N_c g^2}{(4 \pi)^2} \, 
\frac{11}{6} \left( \frac{1}{\epsilon} - \ln \frac{\kappa^2}{\mu^2} + C
\right) \right] \:, \label{gren}
\ee
with a finite constant $C$ given by $(11/6) C = (4/3) C_c + (1/2) C_A
+ C_v$, where $C_v$ is obtained from the finite loop integrals in Eqs.\ 
\eqref{g3ggv1}--\eqref{g3ggv2}.

For the beta function defined in analogy with Eq.\ \eqref{defbeta} we obtain
from Eq.\ \eqref{gren}
\be
\beta_0 = - \frac{11}{3} \, N_c \:,
\ee
which coincides with the one obtained in Eq.\ \eqref{g2beta}
before with the renormalized coupling constant defined through the static 
potential. We should mention that we could have extracted the beta 
function directly from the cutoff dependence of the vacuum wave functional,
as it has actually been done in Ref.\ \cite{MS99} for Yang-Mills theory in 
Weyl gauge. Here, however, our intention was to closely follow the
procedure applied in the Lagrangian covariant formulation.

The integration of the renormalization group equation 
\eqref{defbeta} gives the well-known (one-loop) result
\be
g_R^2 (\kappa) = \frac{(4 \pi)^2}{\ds \frac{11}{3} N_c 
\ln \left( \frac{\kappa^2}{\Lambda_{QCD}^2} \right)} \label{grensol}
\ee
[and the same for $\bar{g}_R^2 (\kappa)$ \eqref{gbar}]. 
It must be noted that for renormalization group improvements like Eq.\
\eqref{grensol} to be sensible we have to suppose that the three-dimensional
formulation presented here is multiplicatively renormalizable to all orders 
in the same way as the usual formulation of a renormalizable covariant quantum 
field theory, which is not known at present (even the renormalizability of the
Lagrangian functional integral approach to Coulomb gauge Yang-Mills theory 
has not yet been shown). Equation \eqref{grensol} and the developments to 
follow are therefore to some degree speculative, but it seemed of some interest
to us to explore the consequences of the natural assumption of multiplicative 
renormalizability.

With these qualifications, we go on to use a standard
renormalization group argument to extract the asymptotic UV behavior of the
equal-time two-point correlation functions. To this end, we differentiate
Eq.\ \eqref{defmultren} with respect to $\kappa^2$ using the 
$\kappa$-independence of the ``bare'' two-point functions. It is then
seen that the $\kappa$-dependence of the renormalized two-point functions
is determined by the anomalous dimensions 
$(\kappa^2 \, \d \ln Z_{A,c}/\d \kappa^2)$. Evaluating the latter
from Eq.\ \eqref{wfr} and replacing $g^2$ in the results with 
$g_R^2 (\kappa)$, we obtain the desired renormalization group equations for 
the equal-time two-point functions, explicitly
\bal
\kappa^2  \frac{\d}{\d \kappa^2} \,
\langle A_{R,i}^a (\bf{p}_1) A_{R,j}^b (\bf{p}_2) \rangle 
&= \frac{N_c g_R^2 (\kappa)}{(4 \pi)^2} \, 
\langle A_{R,i}^a (\bf{p}_1) A_{R,j}^b (\bf{p}_2) \rangle \:, \n \\
\kappa^2  \frac{\d}{\d \kappa^2} 
\langle c_R^a (\bf{p}_1) \bar{c}_R^b (\bf{p}_2) \rangle
&= \frac{4}{3} \, \frac{N_c g_R^2 (\kappa)}{(4 \pi)^2} 
\, \langle c_R^a (\bf{p}_1) \bar{c}_R^b (\bf{p}_2) \rangle \:.
\eal
In these equations, we substitute from Eq.\ \eqref{grensol} for 
$g_R^2 (\kappa)$ and integrate. Using the normalization conditions 
\eqref{rencond} for the determination of the integration constants,
one obtains the momentum dependence of the equal-time two-point functions:
\bal
\langle A_{R,i}^a (\bf{p}_1) A_{R,j}^b (\bf{p}_2) \rangle 
&= \frac{1}{2 |\bf{p}_1|} \left(
\frac{\ds \ln \left( \frac{\kappa^2}{\Lambda_{QCD}^2} \right)}
{\ds \ln \left( \frac{\bf{p}_1^2}{\Lambda_{QCD}^2} \right)} \right)^{3/11}
\hspace{-2mm} \delta^{ab} \, t_{ij} (\bf{p}_1) 
(2 \pi)^3 \delta (\bf{p}_1 + \bf{p}_2) \:, \n \\[2mm]
\langle c_R^a (\bf{p}_1) \bar{c}_R^b (\bf{p}_2) \rangle 
&= \frac{1}{\bf{p}_1^2} \left(
\frac{\ds \ln \left( \frac{\kappa^2}{\Lambda_{QCD}^2} \right)}
{\ds \ln \left( \frac{\bf{p}_1^2}{\Lambda_{QCD}^2} \right)} \right)^{4/11}
\hspace{-2mm} \delta^{ab} \, (2 \pi)^3 \delta (\bf{p}_1 + \bf{p}_2) \:.
\label{2prensol}
\eal
The momentum dependence of the ``bare'' two-point functions, obtained from
Eq.\ \eqref{2prensol} simply by multiplying with the corresponding wave
function renormalization constants $Z_{A,c}$, is obviously the same. By
solving the renormalization group equations for $Z_A$ and $Z_c$ that involve
the anomalous dimensions, it may be shown explicitly that the bare 
two-point functions are $\kappa$-independent, as they must be.

For the static potential, on the other hand, we immediately obtain from
Eqs.\ \eqref{defgbar} and \eqref{grensol} [for $\bar{g}_R^2 (\kappa)$] the 
renormalization group improved result
\be
g^2 V (\bf{p})= \frac{(4 \pi)^2}{\ds \frac{11}{3} N_c \, \bf{p}^2
\ln \left( \frac{\bf{p}^2}{\Lambda_{QCD}^2} \right)} \:.
\ee
Note that this one-loop formula constitutes a very direct expression
of asymptotic freedom.

The result \eqref{2prensol} for the momentum dependence of the equal-time
two-point functions has also been obtained in Ref.\ \cite{Sch08} from a 
Dyson-Schwinger equation for the equal-time ghost correlator, where the
gauge-invariant one-loop running \eqref{grensol} of the renormalized coupling
constant is used as an input. We briefly discuss that derivation here, 
adapted to the conventions of the present paper.

The renormalized equal-time two-point functions are parameterized as
\be
\langle A_{R,i}^a (\bf{p}_1) A_{R,j}^b (\bf{p}_2) \rangle = 
\frac{1}{2\omega(\bf{p}_1^2)} \, \delta^{ab} \, t_{ij} (\bf{p}_1) 
(2 \pi)^3 \delta (\bf{p}_1 + \bf{p}_2)
\ee
and
\be
\langle c_R^a (\bf{p}_1) \bar{c}_R^b (\bf{p}_2) \rangle = 
\frac{d(\bf{p}_1^2)}{\bf{p}_1^2} \, \delta^{ab} \, 
(2 \pi)^3 \delta (\bf{p}_1 + \bf{p}_2) \:,
\ee
and normalized according to the conditions \eqref{rencond}. The
renormalized coupling constant is defined as before in Eq.\ \eqref{grendef}.
Then the Dyson-Schwinger equation for the equal-time ghost two-point function
reads \cite{FR04a}--\cite{ERS07}
\be
d^{-1} (\bf{p}^2) = Z_c - N_c \, g_R^2 (\kappa) \int \frac{d^3 q}{(2 \pi)^3}
\, \frac{1 - (\hat{\bf{p}} \cdot \hat{\bf{q}})^2}{2 \omega (\bf{q}^2)} \,
\frac{d \big( (\bf{p} - \bf{q})^2 \big)}{(\bf{p} - \bf{q})^2} \:. \label{DS}
\ee
Here we have approximated the full ghost-gluon vertex appearing in the
exact equation by the tree-level vertex, as it is appropriate in order to
obtain the (renormalization-group improved) one-loop expressions.

In order to solve Eq.\ \eqref{DS}, we make the following, properly normalized, 
ansatzes for the two-point functions in the ultraviolet limit 
$\bf{p}^2 \gg \Lambda_{QCD}^2$,
\be
\frac{\abs{\bf{p}}}{\omega (\bf{p}^2)} = \left(
\frac{\ds \ln \left( \frac{\kappa^2}{\Lambda_{QCD}^2} \right)}
{\ds \ln \left( \frac{\bf{p}^2}{\Lambda_{QCD}^2} \right)} \right)^\gamma \:,
\qquad d  (\bf{p}^2) = \left(
\frac{\ds \ln \left( \frac{\kappa^2}{\Lambda_{QCD}^2} \right)}
{\ds \ln \left( \frac{\bf{p}^2}{\Lambda_{QCD}^2} \right)} \right)^\delta \:,
\ee
with the exponents $\gamma$ and $\delta$ to be determined. The integral in
Eq.\ \eqref{DS} can then be calculated in the limit 
$\bf{p}^2 \gg \Lambda_{QCD}^2$ and the Dyson-Schwinger equation yields the 
relation \cite{Sch08}
\be
\ln^{-\delta} \left( \frac{\kappa^2}{\Lambda_{QCD}^2} \right)
\ln^{\delta} \left( \frac{\bf{p}^2}{\Lambda_{QCD}^2} \right)
= N_c \, g_R^2 (\kappa) \frac{1}{(4\pi)^2} \frac{4}{3\delta}
\ln^{\gamma+\delta} \left( \frac{\kappa^2}{\Lambda_{QCD}^2} \right)
\ln^{1-\gamma-\delta} \left( \frac{\bf{p}^2}{\Lambda_{QCD}^2} \right) \:,
\ee
from which we infer the sum rule
\be
\label{thelogsumrule}
\gamma + 2 \delta = 1
\ee
for the exponents as well as the identity 
\be
\label{coeffCoulUV}
g_R^2 (\kappa) \frac{1}{(4\pi)^2} \frac{4}{3\delta} N_c 
\ln \left( \frac{\kappa^2}{\Lambda_{QCD}^2} \right) = 1
\ee
for the coefficients. Consistency of the latter relation with the 
well-known perturbative result \eqref{grensol} yields the exponents
\be
\gamma = \frac{3}{11} \;, \qquad \delta = \frac{4}{11} \;,
\ee
where we have used the sum rule \eqref{thelogsumrule} again.
We have thus regained the result of Eq.\ \eqref{2prensol}.

\section{Conclusions}

In this work, we have 
accomplished a systematic perturbative solution of the Yang-Mills
Schr\"o\-din\-ger equation in Coulomb gauge for the vacuum wave functional
following the $e^S$ method in many-body physics. This resulted in
a functional integral representation for
the calculation of equal-time correlation functions. We have derived a
diagrammatical representation of these functions, order
by order in perturbation theory, where the vertices in the diagrams are 
determined from the perturbative calculation of the vacuum wave functional.
The number of the vertices, which by themselves have a perturbative expansion,
grows with the perturbative order. We have determined the equal-time gluon
and ghost two-point correlation functions and the potential between static
color charges to one-loop order in this way.

The results coincide with those of a straightforward calculation in
Rayleigh-Schr\"odinger perturbation theory \cite{CRW09}, and also with the 
values for equal times of the two-point space-time correlation 
functions from a Lagrangian functional integral representation \cite{WR07b}. 
We have emphasized that the latter
coincidence is not trivial since the gauge fixing procedures in the
Hamiltonian and the Lagrangian approach are profoundly different. We
have also used the results of the Lagrangian approach to renormalize
the equal-time two-point correlation functions and the static potential.

With the help of the nonrenormalization of the ghost-gluon vertex which we
also show, or, alternatively, from the static potential, we have extracted
the running of the correspondingly defined renormalized coupling 
constant. The result for the beta function is the one also found in covariant 
and other gauges, $\beta_0 = - (11/3) N_c$ to one-loop order. We have used
standard renormalization group arguments to determine the asymptotic
ultraviolet behavior of the equal-time two-point functions and
the static potential under the assumption of multiplicative renormalizability 
to all orders, with the result that 
\bal
\langle A_i^a (\bf{p}_1) A_j^b (\bf{p}_2) \rangle &\propto 
\frac{\big( \ln (\bf{p}_1^2/\Lambda_{QCD}^2) \big)^{-3/11}}{2 |\bf{p}_1|} \,
\delta^{ab} \, t_{ij} (\bf{p}_1) 
(2 \pi)^3 \delta (\bf{p}_1 + \bf{p}_2) \:, \n \\
\langle c^a (\bf{p}_1) \bar{c}^b (\bf{p}_2) \rangle &\propto
\frac{\big( \ln (\bf{p}_1^2/\Lambda_{QCD}^2) \big)^{-4/11}}{\bf{p}_1^2} \,
\delta^{ab} \, (2 \pi)^3 \delta (\bf{p}_1 + \bf{p}_2) \:, \n \\
g^2 V (\bf{p}_1) &\propto 
\frac{\big( \ln (\bf{p}_1^2/\Lambda_{QCD}^2) \big)^{-1}}{\bf{p}_1^2}
\label{UVres}
\eal
to one-loop order in the perturbative (asymptotically free) regime.

It is clear from the presence of an infinite number of vertices
in the functional integral representation of the equal-time correlation
functions (to infinite perturbative order) that the corresponding 
Dyson-Schwinger equations contain an infinite number of terms, a very
serious problem for the determination of an appropriate approximation scheme 
for nonperturbative solutions. The existence of simplified diagrammatic
rules for the calculation of equal-time correlation functions via the
E-operator, to be appropriately extended to all perturbative orders,
seems to point toward the possibility of formulating similar nonperturbative
equations with a finite number of terms. It would indeed be very interesting
to repeat the type of infrared analysis applied before to Yang-Mills theory
in the Landau gauge \cite{SLR06}, \cite{SHA97}--\cite{PLN04}
and to a variational ansatz in the Coulomb gauge 
\cite{SS01}--\cite{SLR06} for such a set of equations.

The perturbative expression for the vacuum wave functional determined
in this paper, in particular the coefficient functions $f_3$ and $f_4$ in
Eqs.\ \eqref{solf3}--\eqref{solf4ct}, can be used to motivate improved ansatzes
for the vacuum functional going beyond the Gaussian form in a variational 
approach as employed in Refs.\ \cite{SS01}, \cite{FR04a}--\cite{ERS07}. An 
appropriate extension of a Gaussian ansatz would make it possible, e.g.,
to reproduce the correct beta function and the anomalous dimensions of the 
equal-time two-point functions at least to one-loop order in this
approach. This idea is currently being pursued. We note in this context 
that the relation 
\eqref{g2solf2gen} between the coefficient functions $f_2$ and $f_4$ 
following from the Schr\"odinger equation can easily be promoted to an exact 
gap equation, i.e., valid to any 
perturbative order, by replacing $\abs{\bf{p}}$ and $\abs{\bf{q}}$ on the 
right-hand side with $f_2 (\bf{p})$ and $f_2 (\bf{q})$, respectively:
\bal
\lefteqn{\big( f_2 (\bf{p}) \big)^2 \delta^{ab} \delta_{ij}
= \left( \bf{p}^2 - f_2(\bf{p}) \, \frac{N_c g^2}{2}
\int \frac{d^3 q}{(2 \pi)^3} 
\frac{1 - (\hat{\bf{p}} \cdot \hat{\bf{q}})^2}{(\bf{p} - \bf{q})^2} \right) 
\delta^{ab} \delta_{ij}} \hspace{1.5cm} \n \\
&{}+ \frac{1}{2} \int \frac{d^3 q}{(2 \pi)^3} \, 
f_{4;ijkl}^{abcc} (-\bf{p}, \bf{p}, -\bf{q}, \bf{q}) 
t_{kl} (\bf{q}) 
- N_c g^2 \delta^{ab} \int \frac{d^3 q}{(2 \pi)^3} 
\frac{f_2(\bf{p}) - f_2(\bf{q})}{(\bf{p} - \bf{q})^2} \, t_{ij} (\bf{q}) \:.
\eal

Similarly, the explicit perturbative expression for the vacuum wave 
functional in the presence of static external color charges, see Eqs.\ 
\eqref{g2solAAq} and \eqref{g4solAAqq}, is expected to provide valuable 
information in the quest for a detailed understanding of the (quenched) 
interaction between quarks. Different ansatzes for such a wave functional in 
the nonperturbative regime have been considered in Refs.\ \cite{Sch08,SK06}, 
while relevant results from lattice calculations can be 
found in Refs.\ \cite{HIL08,GO09}.

Finally, the results \eqref{UVres} for the ultraviolet behavior of 
the two-point functions are relevant to the corresponding results of a 
numerical evaluation on space-time lattices. A recent numerical calculation
of the equal-time ghost two-point function \cite{NVI09} gives a value of 
$0{.}33(1)$ for the anomalous dimension, quite close to our one-loop result 
$4/11 \approx 0{.}36$. As for the static potential, numerical
results are only available for the instantaneous antiscreening part, given
to one-loop order by Eq.\ \eqref{g4spat}. The asymptotic behavior of this
so-called color Coulomb potential has been determined in Ref.\ \cite{CZ01}
to be
\be
g^2 V_C (\bf{p}_1) \propto 
\frac{\big( \ln (\bf{p}_1^2/\Lambda_{QCD}^2) \big)^{-1}}{\bf{p}_1^2}
\ee
to one-loop order, the same as for the full static potential (except for an
overall factor $12/11$). This ultraviolet behavior was confirmed by earlier
lattice calculations \cite{CZ03}--\cite{VIM07}. However, the most recent 
numerical evaluation \cite{VIM08} contradicts these earlier findings.

The situation is even more controversial for the equal-time (transverse 
spatial) gluon two-point function. Some years ago, scaling violations had
been reported by several groups \cite{VIM07,NTN07,QBC07}. Different proposals 
to understand or deal with the scaling violations have lead to different 
results in the most recent lattice simulations, from zero anomalous dimension 
\cite{BQR09,BQR10} to an anomalous power behavior \cite{NVI09}
\be
\langle A_i^a (\bf{p}_1) A_j^b (\bf{p}_2) \rangle \propto 
\frac{|\bf{p}_1|^{-\eta}}{2 |\bf{p}_1|} \,
\delta^{ab} \, t_{ij} (\bf{p}_1) (2 \pi)^3 \delta (\bf{p}_1 + \bf{p}_2)
\ee
with $\eta = 0{.}40(2)$ (similar to results reported earlier \cite{LM04};
see, however, the corresponding remarks in Ref.\ \cite{QBC07}).
Clearly, a better understanding of the numerical data is needed.
An interesting possibility in this context is the use of anisotropic 
lattices where a drastic reduction of the scaling violations was
reported \cite{NNS09} in the approach to the Hamiltonian limit 
$a_s/a_t \to \infty$ (with the spatial and temporal lattice spacings 
$a_s$ and $a_t$).

Given that the question of multiplicative renormalizability of the 
gluonic two-point correlation function (at equal and at different times)
plays an important r\^ole in the controversy about the scaling violations
\cite{BQR09,NVI09}, it would certainly be interesting to confirm
multiplicative renormalizability in our approach to the two-loop 
level. The extension of the calculations presented here to two loops
appears relatively straightforward, albeit lengthy. Let us mention,
in this context, that the static potential has never been worked out 
explicitly in Coulomb gauge at the two-loop level. In our approach,
the main problem with the two-loop calculations is expected to be the 
correct renormalization of the expressions. Corresponding results for 
the space-time correlation functions from a Lagrangian functional
integral approach are not known to this level, so one cannot proceed
in analogy with Section 4. The calculations with the Lagrangian functional 
integral method are complicated, at the two-loop level, by the appearance
of Christ-Lee-Schwinger terms \cite{CL80} (see also Refs.\ \cite{CT86,Dou87})
which are required in order to produce results corresponding to the 
Hamiltonian \eqref{christlee} with its specific operator ordering 
(in particular, the insertion of powers of the Faddeev-Popov determinant). 
As an alternative to the procedure of Section 4, one may consider the use
of a simple three-dimensional ultraviolet momentum cutoff. Care has
to be taken, however, since such a cutoff can break
the covariance of the theory (see Ref.\ \cite{Web09b} for an example
in Yukawa theory), in which case appropriate noncovariant
counterterms have to be included.

\subsection*{Acknowledgments}

It is a pleasure to thank Adam Szczepaniak and 
Peter Watson for many valuable discussions on the Coulomb gauge.
A.W. is grateful to the Institute for Theoretical Physics at the University of
T\"ubingen for the warm hospitality extended to him during a two-months stay 
in the summer of 2008. Support by the Deutscher Akademischer Austauschdienst
(DAAD), Conacyt grant 46513-F, CIC-UMSNH,
Deutsche Forschungsgemeinschaft (DFG) under contract Re 856/6-3, and 
Cusanuswerk--Bisch\"ofliche Studienf\"orderung
is gratefully acknowledged.

\begin{appendix}

\section{Gauge transformations in Lagrangian and Hamiltonian formalisms}

While the Lagrangian approach to Yang--Mills theory offers some convenient
features (such as manifestation of Lorentz invariance), the more cumbersome
Hamiltonian approach yields equations of motion invariant under a larger set
of gauge transformations. In what follows, we discuss gauge invariance
starting from the classical Lagrangian and Hamiltonian functions, 
respectively, prior to quantization. We will employ standard covariant notation
in this appendix; in particular, spatial subindices refer to the covariant
components of the corresponding 4-vector or tensor.

The Lagrangian function of the gauge sector,
\be
L=-\frac{1}{4}\int d^3x\: F_{\mu\nu}^a(x)F^{\mu\nu}_a(x)\; ,
\ee
is invariant under gauge transformations of the gauge field $A_\mu(x)\equiv A_\mu^a (x)T^a$,
\be
\label{Ltra}
A_\mu(x)\rightarrow U(x)A_\mu(x)U^\dagger(x) + \frac{1}{g}U(x)\partial_\mu U^\dagger(x) \; ,
\ee
where $U\in SU(N)$ and $[T^a,T^b]=f^{abc}T^c$.

The Weyl gauge, $A_0^a(x)=0$, can be found by choosing the time-ordered exponential
\be
\label{Weyl}
U^\dagger(x)= \textrm{T}\:\exp\left( - g \int^t dt' A_0(\bf{x},t') \right) \; .
\ee
To remain in the Weyl gauge, the transformation \eqref{Weyl} may be followed by
time-independent transformations $U(\bf{x})$ only. We can therefore fix the Coulomb
gauge, $\partial_i A^i_a(x)=0$, at one instant of time but it is impossible to fix
both gauges simultaneously for all times.

In the Hamiltonian formalism, on the other hand, gauge transformations are generated
by (first-class) constraints in configuration space \cite{Dir64}. To see that,
supplement the Hamiltonian function 
\be
H=\frac{1}{2}\int d^3x\: \left(\bf{\Pi}_a^2(x) + \bf{B}_a^2(x)\right) - \int d^3 x \: A_0^a(x) \hat D_i^{ab}(x)\Pi_b^i(x)
\ee
by the constraints
\be
\label{phis}
\phi_1^a(x)=\Pi_0^a(x)\approx 0\; ,\quad \phi_2^a(x)=\hat D^{ab}_i(x)\Pi^i_b(x)\approx 0\;
\ee
with some arbitrary Lagrange multiplier fields $\{\lambda_k^a(x)\}$,
\be
\label{H_E}
H_E=H+\sum_{k=1,2}\int d^3 x \:\lambda_k^a(x)\phi_k^a(x)\; .
\ee
We defined $\Pi_\mu^a(x)=F_{\mu 0}^a(x)$ and
$\hat D^{ab}_i(x)=\delta^{ab}\partial_i-gf^{abc}A_i^c(x)$.
The extended
Hamiltonian $H_E$ in Eq.\ \eqref{H_E} is equivalent to the original
Hamiltonian $H$ since the constraints $\{\phi_k^a(x)\}$ vanish weakly
(in the Dirac sense \cite{Dir64}). The infinitesimal time evolution of
the gauge field $A_\mu^a({\bf{x}},t)$ from $t_0$ to $t=t_0+\delta t$,
generated by $H_E$ through the Poisson brackets,
\be
A_\mu^a({\bf{x}},t)= A_\mu^a({\bf{x}},t_0)+\delta t \:\{ A_\mu^a({\bf{x}},t_0),H\} + \delta t \sum_{k=1,2}\int d^3 y \: \lambda_k^b(y) \:\{ A_\mu^a({\bf{x}},t_0), \phi_k^b(y)\} \; ,
\ee
gives for two different sets of Lagrange multiplier functions
$\{\lambda_k'^b(x)\}$ and $\{\lambda_k''^b(x)\}$ two different
results $A_\mu'^a$ and $A_\mu''^a$, respectively. These differ to 
${\cal{O}}( \delta t )$ by
\be
\label{fdiff}
A_\mu''^a({\bf{x}},t) - A_\mu'^a({\bf{x}},t) = \delta t \sum_{k=1,2} \int d^3y \:\left(\lambda_k''^b(y)-\lambda_k'^b(y)\right) \{ A_\mu^a({\bf{x}},t), \phi_k^b(y)\} 
\ee
and are physically equivalent. Thus, the function
\be
G=\sum_{k=1,2} \int d^3y \:\tau_k^a(y)\phi_k^a(y)
\ee
generates infinitesimal gauge transformations in the (extended) Hamiltonian
formalism with arbitrary functions $\tau_1^a(x)$ and $\tau_2^a(x)$. Computing
the Poisson brackets in Eq.\ \eqref{fdiff} yields
\bal
\label{A0trafo}
A_0^a(x) & \rightarrow A_0^a(x) + \tau_1^a(x) \\
\label{Aktrafo}
A_i^a(x) &\rightarrow A_i^a(x) - \hat D^{ab}_i (x)\tau_2^b(x)
\eal
The difference to the gauge transformations \eqref{Ltra} in the Lagrangian
formalism is that the time component and the spatial components of the gauge
field transform independently. The two functions $\tau_1^a(x)$ and $\tau_2^a(x)$
allow for a larger set of gauge transformations than the single function $U(x)$
in the Lagrangian formalism. The simultaneous fixing of Weyl and Coulomb gauges,
which is impossible in the Lagrangian formalism, can be accomplished in the
Hamiltonian formalism by appropriately choosing $\tau_1^a(x)$ and $\tau_2^a(x)$
(see Ref.\ \cite{Cos84} for the abelian case). Subsequently, the non-abelian
gauge-fixed theory can be canonically quantized with projection on the physical
Hilbert space \cite{CL80}, or with Dirac brackets \cite{Sch08} enforcing all
constraints strongly. Both quantization prescriptions produce the Hamiltonian
operator given by Eq.\ \eqref{christlee}.

\end{appendix}

\end{document}